\documentclass[sn-chicago]{sn-jnl}

\usepackage{graphicx}%
\usepackage{multirow}%
\usepackage{amsmath,amssymb,amsfonts}%
\usepackage{amsthm}%
\usepackage{mathrsfs}%
\usepackage[title]{appendix}%
\usepackage{xcolor}%
\usepackage{textcomp}%
\usepackage{manyfoot}%
\usepackage{booktabs}%
\usepackage{algorithm}%
\usepackage{algorithmicx}%
\usepackage{algpseudocode}%
\usepackage{listings}%
\usepackage[latin1]{inputenc}
\usepackage{graphicx}
\usepackage{color}
\usepackage{geometry}  
\usepackage{parskip}   
\usepackage{amsmath, amssymb} 
\usepackage{setspace} 
\usepackage{booktabs} 
\DeclareMathOperator{\E}{\mathbb{E}}
\usepackage{float}
\usepackage{tabularx}
\usepackage{subcaption}
\usepackage{lscape}
\usepackage{afterpage}
\usepackage{graphicx}
\usepackage{orcidlink}

\theoremstyle{thmstyleone}%
\theoremstyle{thmstyletwo}%

\theoremstyle{thmstylethree}%

\newcommand{\bs}[1]{\ensuremath{\boldsymbol{#1}}} 

\newcolumntype{Y}{>{\centering\arraybackslash}X}

\begin{document}

\title[Flexible tree-structured regression for clustered data]{Flexible tree-structured regression for clustered data with an application to quality of life in older adults}


\author*[1]{\fnm{Nikolai} \sur{Spuck}\orcidlink{0000-0001-6345-9634}}\email{spuck@imbie.uni-bonn.de}

\author[1]{\fnm{Matthias} \sur{Schmid}\orcidlink{0000-0002-0788-0317}}

\author[1]{\fnm{Moritz} \sur{Berger}\orcidlink{0000-0002-0656-5286}}

\affil[1]{\orgdiv{Institute of Medical Biometry, Informatics, and Epidemiology}, \orgname{Medical Faculty, University of Bonn}, \orgaddress{\street{Venusberg-Campus 1}, \city{Bonn}, \postcode{53127}, \country{Germany}}}


\abstract{Tree-structured models are a powerful alternative to parametric regression models if non-linear effects and interactions are present in the data. Yet, classical tree-structured models might not be appropriate if data comes in clusters of units, which requires taking the dependence of observations into account. This is, for example, the case in cross-national studies, as presented here, where country-specific effects should not be neglected. To address this issue, we present a flexible tree-structured approach that achieves a sparse modeling of unit-specific effects and identifies subgroups (based on individual-level covariates) that differ with regard to the outcome. The methodological advances were motivated by the analysis of quality of life in older adults using data from the survey of Health, Ageing and Retirement in Europe. Application of the proposed model yields promising results and illustrated the accessibility of the approach. A comparison to alternative methods with regard to variable selection and goodness-of-fit was performed in several simulation experiments.}

\keywords{\vspace{-0.5cm}CASP score, clustered data, tree-based models, tree-structured clustering}

\pacs[Funding]{Support by the German Research Foundation (DFG), grant BE 7543/1-1, is gratefully acknowledged.}

\maketitle

\section{Introduction}

People's quality of life (QoL) is essential in evaluating and guiding many health, social, community and environmental policy actions \citep{Bowling2011}. Often, QoL is of particular interest in the group of older adults since they tend to make up a larger proportion of the population in most industrialized countries each year and are most likely to experience events that negatively affect their autonomy and everyday life \citep{BorratBesson2015}. According to Eurostat \citep{Eurostat2024} the median age of the population in the EU increased from 39.0 years in 2003 to 44.5 years in 2023. In order to provide an explicit and well-defined measure for QoL in older adults, \citet{Hyde2003} developed the so-called Control, Autonomy, Self-Realization and Pleasure (CASP) scale comprising 19 Likert-type items on these four domains. The CASP-19 scale has become a widely applied and well-established tool in studies investigating QoL in older adults, see, among others, \citet{Sim2011}, \citet{Howel2012}, \citet{Kim2015}, and \citet{FriasGoytia2024}.  

Here, we analyze data from the survey of Health, Ageing and Retirement in Europe, in short SHARE (see \citealp{BoerschSupan2013}, for methodological details). The main objective of SHARE is to collect panel data that enables researchers to investigate the impact of socio-economic and health-related factors on the ageing process. Moreover, SHARE constitutes a cross-national survey that is aimed to explore the differences between European countries in dealing with the consequences of population ageing. SHARE provides information on individuals aged 50 years and older gathered in 27 European countries and Israel. QoL was measured on a SHARE-specific CASP scale, which utilizes an adapted 12-item version of the CASP-19 questionnaire \citep{BorratBesson2015}. 

When analyzing QoL in SHARE one has to deal with the issue that the data is clustered by country and therefore observations can not be treated as independent. It appears sensible to assume that measurements within units (here countries) tend to be more similar than measurements between units. This heterogeneity needs to be taken into account using an appropriate regression approach. In our application we consider a sample of $n=45,038$ observations from the ninth wave of SHARE collected from October 2021 to October 2022 \citep{Bergmann2024, Share2024}, which contains between $391$ (Israel) and $3,116$ (Belgium) observations per country. In this paper, we propose a novel approach for modeling the CASP score using a tree methodology that (i) accounts for heterogeneity between the 28 different countries by sparse modeling of country-specific effects, and (ii) is able to identify distinct subgroups of individuals which differ with regard to their CASP score based on socio-economic and health-related factors as well as their interactions.  

Regression approaches for modeling heterogeneity among units are manifold. The most popular tool is \textit{mixed effects regression}, for example, in SHARE a model with country-specific random intercepts. Mixed effects regression models postulate that the random effects follow a common predefined distribution (typically a normal distribution), which results in a parsimonious model specification \citep{Verbeke2000, Molenberghs2005}. This strong assumption, however, comes at the price that statistical inference may be sensitive to a misspecification of the random effects distribution \citep{Heagerty2001, Litiere2007}. In addition, \citet{Grilli2011} showed that a correlation between random effects and explanatory variables may lead to biased effect estimates. An alternative to mixed models are \textit{fixed effects models}, in which each country has its own parameter. In the literature fixed effects models are also referred to as ``no-pooling'' models \citep{Gelman2007} and are based on the assumption that the country-specific effects are unrelated and exist completely independently from each other \citep{Bell2018}. 

To overcome both the limitations of mixed and fixed effects models, it can alternatively be assumed that the unit-specific effects follow a more flexible discrete distribution. This implies that there are groups of units sharing the same effect. In our application, the identification of groups of countries that are similar with regard to their QoL and the interpretation of relevant differences are of great interest. Clustering of units can be achieved by using finite mixtures of regression models \citep{Gruen2007}, by Bayesian mixed models with Dirichlet process prior \citep{Heinzl2013} and within fixed effects models applying penalized maximum likelihood estimation \citep{Oelker2017} or tree-based splits \citep{Berger2018}. The latter, which we are focusing on here, is based on a fixed effects model containing tree-structured unit-specific intercepts and a linear function of a set of explanatory variables. \citet{Berger2018} demonstrate that their approach is very flexible in capturing heterogeneity among units particularly in scenarios where the distribution of random effects is skewed and in scenarios with correlation between random effects and covariates. Yet, the approach by \citet{Berger2018} is still limited as it only uses a linear combination of the explanatory variables in the predictor function. When modeling associations in SHARE the assumption of linearity may be too restrictive as it does not account for possible non-linear effects and interactions between socio-economic and health-related factors of interest (for example, level of income and chronic diseases). To address this issue, we propose a regression model extending the approach by \citet{Berger2018} that comprises two tree structures: One tree determining unit-specific (country-specific) effects, and one tree modeling the effects of covariates (individual-level health-related and socio-economic factors). 

The underlying concept of \textit{recursive partitioning} or \textit{tree-based modeling} originates from the framework of classification and regression trees (CART) proposed by \citet{Breiman1984}. When growing a classical tree the predictor space is partitioned into a set of disjoint subsets by sequentially applying binary splits. In each subset a simple model (for example, a constant) is fitted. Overviews and comparisons of recursive partitioning methods have been given by \citet{Strobl2009}, \citet{Doove2014} and \citet{Kern2018}. The tree methodology applied here (see Section \ref{sec:fitting} for a detailed description of the algorithm) slightly differs from theses approaches, as we do not apply a traditional recursive partitioning algorithm, but fitting and tree building is performed within the framework of fixed effects models. The key advantages of our proposed model are (i) the flexibility in capturing the effects of individual-level factors (including non-linear effects and interactions), (ii) its built-in mechanism to select the relevant factors, and (iii) sparse modeling of unit-specific effects assuming a discrete distribution. 

The remainder of this article is structured as follows: In Section \ref{sec:modeling} we introduce the notation, describe the proposed tree-structured model and discuss alternatives based on random effects. Details of the fitting procedure are outlined in Section \ref{sec:fitting}. In Section~\ref{sec:app} we apply the proposed model for analyzing the CASP score in the SHARE data. In Section~\ref{sec:simulation} the proposed model is compared to alternative methods based on several simulation experiments. The article concludes with a summary and discussion on the different methods for modeling heterogeneity (Section~\ref{sec:disc}). 

\section{Regression for clustered data}\label{sec:modeling}

Consider clustered data with $n$ units given by $(y_{ij}, \bs{x}_{ij}),\, i = 1,\dots, n,\,  j= 1\dots, n_i$, where~$y_{ij}$ denotes the value of the outcome variable of observation~$j$ from unit~$i$ and $\bs{x}_{ij}^\top = (x_{ij1},\dots x_{ijp})$ denotes the vector of a set of covariates. In general, it is assumed that the values of the covariates vary within units and that the number of observations per unit $n_i$ may differ across units. In the following, alternative parametric and non-parametric approaches for modeling clustered data are considered. The focus is mainly on models with unit-specific intercepts. 

\subsection{Models with random effects}
\label{subsec:RE}
 
In classical \textit{generalized linear mixed effects models} (GLMMs; \citealp{Verbeke2000}) with random intercepts, the expectation of the outcome variable $\mu_{ij}=\E (y_{ij}|\bs{x}_{ij},b_i)$ is linked to the covariates in the form 
\begin{equation}
\label{glmm}
g\left(\mu_{ij}\right) = \eta (\bs{x}_{ij}, b_i ) = \beta_{0} +  \bs{x}_{ij}^\top\,\bs{\beta} + b_i\, ,
\end{equation}
where $g(\cdot)$ denotes a suitable link function, $\bs{\beta}$ is the vector of regression coefficients (that is, the vector of fixed effects of the covariates) and $b_i$ denotes the random intercept of unit $i$. It is commonly assumed that the random intercepts follow a normal distribution, i.e. $b_i\sim N(0, \sigma_b^2)$. This distributional assumption on the random intercepts makes the GLMM very efficient, as only the variance parameter has to be estimated in the random effects part of the model. 

The simple form of the GLMM in Equation~\eqref{glmm} comes with the drawback that only linear main effects of the covariates on the outcome are assumed. This, however, may be too restrictive in real-world data (for example, in our application to SHARE), as it does not account for possible non-linear effects and interactions between covariates. To address this issue \citet{Hajjem2011} and \citet{Sela2012} simultaneously proposed a flexible non-parametric approach using recursive partitioning. Their approaches, referred to as mixed effects regression trees (MERT) and RE-EM trees, respectively, combine a simple random intercept model with a standard regression tree. The predictor function of a RE-EM tree can be written as
\begin{equation}
\label{reem}
\eta (\bs{x}_{ij}, b_i) = \beta_0 + tr(\bs{x}_{ij} )+ b_i\, ,
\end{equation} 
where the function $tr(\cdot)$ is determined by a tree structure. This means that $tr(\cdot)$ sequentially partitions the observations into disjoint subsets $N_m,m = 1,\dots,M$, based on the values of the covariates and assigns a constant $\gamma_m$ to each subset~$N_m$ (by averaging the respective outcome values). The constant $\gamma_m$ can also be interpreted as the regression coefficient in $N_m$. Hence, the function $tr(\cdot)$ is given by
\begin{equation}
\label{tr}
tr(\bs{x}_{ij}) = \sum_{m = 1}^{M} \gamma_{m}\,I(\bs{x}_{ij} \in N_m )\, ,
\end{equation}
where $I(\cdot)$ denotes the indicator function. Importantly, higher-order interactions can be captured by the tree in a very flexible way. RE-EM trees are fitted iteratively by alternating between two steps: (i) Fitting the tree $tr(\bs{x}_{ij})$ by applying the CART algorithm, while keeping the random effects fixed, and (ii) estimating the random intercepts, while keeping the tree structure fixed. More recently, \citet{Fu2015} introduced an adapted version of the RE-EM tree that applies conditional inference trees instead of CART. In a similar vein, an flexible tree-based approach building on the framework of conditional inference trees has been proposed by \citet{Fokkema2018}. 

Both, GLMMs and RE-EM trees specify normally-distributed random intercepts to describe the heterogeneity between units. This is useful if the focus mainly is on the effects of the covariates (particularly, in scenarios, where $n \ggg n_i$). Yet, in our analysis of the CASP score in SHARE, we are explicitly interested in cross-national differences, that is, in the country-specific effects. We therefore propose to use a fixed effects model instead, which is outlined in the next section.   
 
\subsection{Models with tree-structured fixed effects}
\label{subsec:TSC}

An alternative to the mixed effects models introduced in the previous section, are \textit{fixed effects models} (FEMs) with predictor function 
\begin{equation}
\label{fixed}
\eta (\bs{x}_{ij}, \beta_{0i} ) = \beta_{0i} + \bs{x}_{ij}^\top\,\bs{\beta}\, ,
\end{equation}
where each unit has its own parameter $\beta_{0i}$. The specification of one parameter per unit can easily turn into problems, because it results in a very large number of coefficients, which affects estimation accuracy and complicates the interpretation of effects. For example, in wave 9 of SHARE 28 country-specific intercepts need to be estimated when using the FEM in \eqref{fixed}. To deal with this issue, \citet{Berger2018} proposed a tree-structured FEM, which assumes that there are groups of units that share the same effect on the outcome variable. Building clusters of units highly reduces the number of parameters and increases stability of the estimates. The tree-structured FEM by \citet{Berger2018} has the form
\begin{equation}
\label{ltsc}
\eta(\bs{x}_{ij}) =  tr_{\text{0}}(i) + \bs{x}_{ij}^\top\,\bs{\beta}\, ,
\end{equation}
where $tr_{\text{0}}(\cdot )$ describes the unit-specific intercepts represented by a tree structure. The tree forms clusters of units with the same effect on the outcome and is given by 
\begin{equation}
\label{trH}
tr_{0} (i) = \sum_{c=1}^{C} \beta_{0c}\,I(i\in N_{0c} )\, ,
\end{equation} 
where $C$ denotes the number of identified clusters $N_{0c}$ and $\beta_{0c}$ is the corresponding cluster-specific intercept. To obtain $tr_{0}(i)$,  the observations are sequentially partitioned into disjoint subsets using the unit number as the only covariate, while the other parameters (effects of the covariates) are fitted simultaneously to all observations. \citet{Berger2018} proposed to treat the unit number as ordinal variable by ordering the units with respect to their means of the outcome variable before tree building. 

Just like the GLMM, the tree-structured FEM in Equation~\eqref{ltsc} is restricted to linear main effects of the covariates, only. To overcome this limitation and inspired by the works of \citet{Hajjem2011} and \citet{Sela2012} on mixed effects models, we propose a tree-structured FEM, where the effects of the covariates are also determined by a tree structure. Specifically, the predictor function of our proposed model contains two trees and can be written as 
\begin{equation}
\label{ttsc}
\eta(\bs{x}_{ij}) =  tr_{0}(i) + tr(\bs{x}_{ij})\, ,
\end{equation}
where $tr(\cdot)$ and $tr_{0}(\cdot )$ are defined as in Equations~\eqref{tr} and \eqref{trH}, respectively.
The model in~\eqref{ttsc} is constructed in a stepwise procedure, where in each step either a split in the tree of the covariates $tr(\cdot)$ or in the tree that determines the clustering of units $tr_{0}(\cdot)$ is performed. The starting point is a simple model with a global intercept, only. Assuming that a split in $x_k$ at split point $c_k$ is selected in the first step, results in a model with predictor function
\begin{equation}
\eta^{[1]}(\bs{x}_{ij}) = \beta_0^{[1]} +  \gamma_{1}^{[1]}I(x_{ijk} \leq c_k)\, ,
\end{equation}
where $\beta_0^{[1]}$ is a global intercept and $\gamma_{1}^{[1]}$ is the effect on the outcome in the left node. Note that the right node $\{x_{ijk} > c_k\}$ in $tr(\cdot)$ serves as a reference node to ensure parameter identifiability. In the second step, either one of the current nodes is split further or a split with regard to the one of the units in the intercept tree is performed. Let us assume splitting the units into the clusters $N_{01}$ and $N_{02}$ is the second step. This yields the predictor function  
\begin{equation}
\eta^{[2]}(\bs{x}_{ij}) = \beta_{01}^{[2]}I(i\in N_{01}) + \beta_{02}^{[2]}I(i\in N_{02}) + \gamma_{1}^{[2]}I(x_{ijk} \leq c_k)  \, ,
\end{equation}
where $\beta_{01}^{[2]}$ and $\beta_{02}^{[2]}$ are the unit-specific intercepts in the two selected nodes and $\gamma_{1}^{[2]}$ is an update of the parameter from the previous iteration. To determine the split in $tr_{0}(\cdot)$ the unit number is treated as an ordinal variable (see Section~\ref{sec:fitting} for details on the fitting procedure). A third split in $tr(\cdot)$ with regard to $x_\ell$ at split point $c_\ell$ in the left node then results in a predictor of the form 
\begin{align}
\eta^{[3]}(\bs{x}_{ij}) = &\beta_{01}^{[3]}I(i\in N_{01}) + \beta_{02}^{[3]}I(i\in N_{02})\nonumber \\
& + \gamma_{1}^{[3]}I(x_{ijk} \leq c_k)I(x_{ij\ell} \leq c_\ell) + \gamma_{2}^{[3]}I(x_{ijk} \leq c_k)I(x_{ij\ell} > c_\ell)   \, ,
\end{align}
with the new effects $\gamma_{1}^{[3]}$ and $\gamma_{2}^{[3]}$. 

It is important to note that the coefficients of the tree-structured model in~\eqref{ttsc} can only interpreted with regard to a reference node. For example, if the outcome variable $y_i$ is metrically scaled and $g(\cdot)$ is the identity link, the coefficient $\beta_{01}^{[3]}$ denotes the expected values of the outcome variable in cluster $N_{01}$ given that $x_{ijk} > c_k$ (that is, for the subgroup in the reference node). Analogously, the coefficients $\gamma_{1}^{[3]}$ and $\gamma_{2}^{[3]}$ determine the effects on the outcome variable compared to the subgroup in the reference node. To allow for a more intuitive interpretation of the model coefficients, we propose to apply the adjustment 
\begin{align}  
\label{adj}
&\tilde{\beta}_{0c} = \beta_{0c} + \bar{\gamma} \quad \text{and} \nonumber\\
&\tilde{\gamma}_{m} = \gamma_{m} - \bar{\gamma}\, ,
\end{align} 
where $\bar{\gamma}=\frac{1}{n}\sum_{i = 1}^{n}\frac{1}{n_i}\sum_{j = 1}^{n_i} tr(\bs{x}_{ij})$ denotes the mean of the coefficients in the tree of the covariates across all individuals. The coefficients $\tilde{\beta}_{0c}$ can then be interpreted as the \textit{average cluster-specific intercepts} and the coefficients $\tilde{\gamma}_{m}$ represent subgroup-specific effects compared to these averages. In case of a metrically scaled outcome variable (see also our application to SHARE in Section~\ref{sec:app}) this translates into the expected values for each cluster ($\tilde{\beta}_{0c}$) and subgroup-specific deviations from these expectations ($\tilde{\gamma}_{m}$). More details on the fitting procedure are given in the next section.

\section{Fitting procedure}\label{sec:fitting}

In each step of the tree-building algorithm, the best split among all candidate variables (that is, one component of $\bs{x}$ or the unit number $i$) and among all possible split points is selected, starting from a predictor function with global intercept, only. For this, all possible models with one additional split in either the tree of the covariates $tr(\cdot)$ or the tree that determines the clustering of units $tr_{0}(\cdot)$ are evaluated and the best performing one yielding the smallest deviance is selected. In FEMs the deviance is a quite natural measure of the model fit. This criterion is also equivalent to minimizing the entropy, which has been used as a splitting criterion already in the early days of tree construction \citep{Breiman1984}. Note that, in contrast to common trees, in each step of the algorithm all the observations are used to derive estimates of the model parameters. Hence, all parameters are refitted in each iteration and no previously estimated parameters are kept. This ensures that one obtains valid estimates of the two tree components (the coefficient estimates of either of the two components are adjusted for the change through a split in the other) together with the splitting rule.

When selecting the first split in $tr_{0}(\cdot)$ with regard to the unit number, one has to consider $2^{n-1}$ possible partitions, which may be a very large number. To avoid this exponential computational cost, we instead order the units with respect to their means of the outcome variable $\bar{y}_i$ beforehand and treat the unit number as an ordinal variable during tree building. Therefore, only $n - 1$ possible splits have to be considered. This approach, which has also been used by \citet{Berger2018}, has been shown to work well in earlier research, see \citet{Breiman1984} and \citet{Ripley1996} for binary outcomes and \citet{Fisher1958} for continuous outcomes.

To determine the optimal number of splits and hence the size of the trees, we use a post-pruning strategy, where a large number of splits $S_{\text{max}}$ is carried out first and afterwards the trees are pruned to an adequate size to prevent overfitting. Running the stepwise algorithm (with a sufficiently large number of splits) results in a sequence of nested models. These models can be evaluated with regard to their goodness of fit applying a likelihood-based criterion. Specifically, we suggest to select the optimal number of splits by maximizing the cross-validated predictive log-likelihood. In the simulation study and our application to SHARE, we use $10$-fold cross-validation and additionally apply the one standard error (1SE) rule. That is, one selects the model yielding a cross-validated log-likelihood value within one standard error of the model with the maximal value. This is in line with the algorithm by \citet{Sela2012}. Subsequently, the final model with the selected number of splits is fitted to the entire data.

To prevent the algorithm from building extremely small nodes (with only a few observations), an additional \textit{minimal bucket size} constraint $n_{\text{mb}}$  may be applied. With the minimal bucket size constraint, the minimum number of observations required in any terminal node is limited downward. 

To summarize, the following steps are performed during the fitting procedure:
\begin{enumerate}
\item[1.] \textbf{Ordering of units:} Order the units $i\in \{ 1,\dots, n\}$ according to the average values of the outcome variable in each unit $\bar{y}_i$ and initialize the corresponding ordinal variable.
\item[2.] \textbf{Initial model:} Fit the model without any covariates, yielding a single estimate of the intercept $\hat{\beta}_0$.
\item[3.] \textbf{Tree building:} Set $s=1$.
\begin{itemize}
\item[(a)] Fit all candidate models with one additional split regarding one of the covariates or the unit $i$, that fulfill the minimal bucket size constraint, in one of the already built nodes. If none of the additional splits meets the minimal bucket size constraint, terminate the algorithm.
\item[(b)] Select the best performing model based on the minimal deviance.
\item[(c)] Fit the selected model and set $s=s+1$. If $s<S_{\text{max}}$, continue with step (3a). 
\end{itemize}
\item[4.] \textbf{Post-pruning:} Select the optimal model from the sequence of nested models generated in steps (2) and (3) based on the predictive log-likelihood applying $k$-fold cross-validation with the 1SE rule. Then fit the model with the corresponding number of splits to the complete data set.
\end{enumerate}

Technically, the proposed algorithm can be embedded into the framework of tree-structured varying coefficients models (TSVC; \citealp{Berger2019}). The models can therefore be fitted by the eponymous \textsf{R} add-on package \textbf{TSVC} \citep{Berger2018TSVC}, where the covariates $\bs{x}$ and the unit number $i$ serve as effect modifiers, modifying the effect of constant auxiliary variables.

\section{Application: Quality of life in SHARE}
\label{sec:app}

SHARE is a longitudinal, cross-national survey that collects data from individuals aged 50 years and older living in the European Union and Israel \citep{BoerschSupan2013}. Data collection for the first wave of SHARE started in 2004 in 19 different countries. Since then a total of nine waves have been conducted. The survey was mainly designed to provide information on how socio-economic and health-related factors influence the aging process. Here, we analyze data from the ninth wave collected from October 2021 to October 2022 across 28 countries \citep{Bergmann2024, Share2024}. The objective of our analysis was the flexible modeling of QoL in terms of the CASP score by (i) accounting for country-specific effects in a sparse way, and (ii) identifying subgroups of individuals which differ in their CASP score based on socio-economic and health-related factors. 

\begin{figure}[!t]
\centering
\includegraphics[width = 0.7\textwidth]{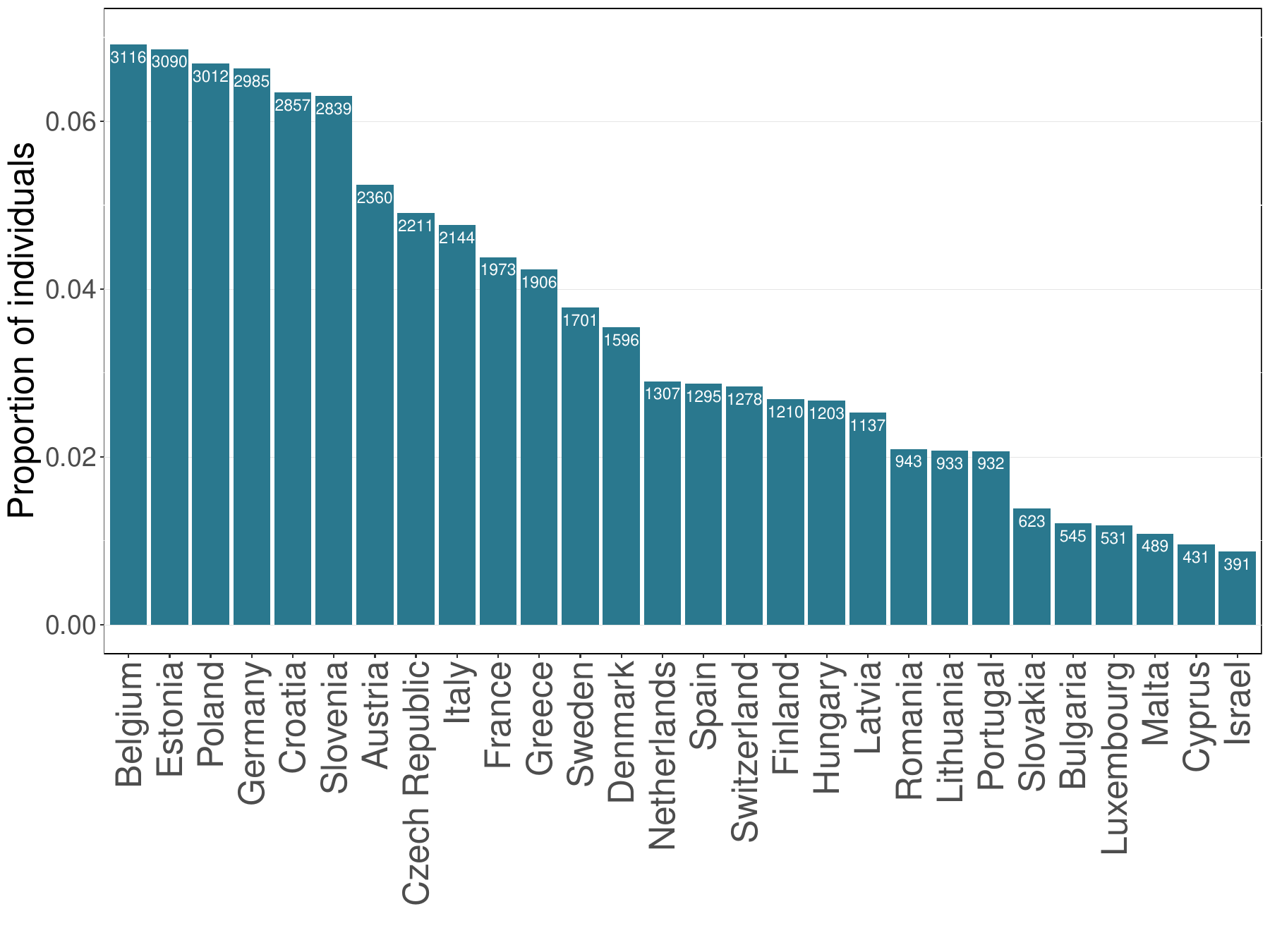}
\caption{Analysis of the SHARE data: Distribution of individuals by country. Absolute and relative frequencies of individuals per country included in the analysis data set}\label{share_countries}
\end{figure}

In a preliminary step, for households with more than one individual participating in the survey one representative was selected at random. This resulted in an analysis data set of  $n = 45, 038$ individuals from 28 countries. Figure~\ref{share_countries} shows the distribution of individuals included in the analysis by country. The country with the largest number of participants was Belgium with $n = 3, 116$, whereas only $n = 391$ participants from Israel were eligible for our analysis (which constitutes the lowest number of participants). The individual-level factors considered for modeling were: sex, age (in years), number of people living in the household, number of children, number of chronic diseases, educational level, employment status, and the level of income (the income decile which the household falls in by country). Summary statistics of these factors are given in Table~\ref{tab:summary}. For more details on the ninth wave of SHARE, see also \citet{Bergmann2024} and \citet{Share2024}.

\begin{table*}[!t]
\caption{Analysis of the SHARE data. Summary statistics of the individual-level factors included in the analysis}\label{tab:summary}
\begin{center}
\begin{small}
\begin{tabularx}{\textwidth}{l c *{6}{Y}}
\toprule
\bf{Variable}& \multicolumn{6}{c}{\bf{Summary statistics}}\\
\midrule
& $x_{min}$ & $x_{0.25}$ & $x_{med}$ & $\overline{x}$ &$x_{0.75}$ & $x_{max}$ \\
Age &  50 & 62  & 69  & 69.3 & 76 & 105 \\
Household size & \ \ 1 & \ \ 1 & \ \ 2 & \ \ 1.9 & \ \ 2 & \ \ 11 \\
Number of children & \ \ 0 & \ \ 1 & \ \ 2 & \ \ 2.0 & \ \ 3 & \ \ 17 \\
Number of chronic diseases & \ \ 0 & \ \ 1 & \ \ 2 & \ \ 1.9 & \ \ 3 & \ \ 14 \\
\midrule
\end{tabularx}
\begin{tabularx}{\textwidth}{l l Y r}
Sex & Male (0) & & $18\, 166\, (40.3\%) $ \\
& Female (1) & & $26\, 872\, (59.7\%)$ \\
Education & Pre-education (0) & & $\ \ 1\, 123\, (\ \ 2.5\%)$ \\
& Primary education (1) & & $\ \ 5\, 242\, (11.6\%) $ \\
& Secondary education first stage (2) & & $\ \ 7\, 214\, (16.0\%)$ \\
& Secondary education second stage (3) & & $17\, 919\, (39.8\%)$ \\
& Post-secondary education (4) & & $\ \ 2\, 295\, (\ \ 5.1\%)$ \\
& Tertiary education first stage (5) & & $10\, 864\, (24.1\%) $ \\
& Tertiary education second stage (6) & & $\ \ \ \ \, \ 391\, (\ \ 0.9\%)$ \\
Employment & Unemployed or retired (0) & & $35\, 197\, (78.1\%)$ \\
& Employed or self employed (1) & & $\ \ 9\, 841\, (21.9\%)$ \\
\bottomrule
\end{tabularx}
\end{small}
\end{center}
\end{table*}

We fitted the proposed tree-structured FEM~\eqref{ttsc} to the analysis data set, where the socio-economic and health-related factors presented in Table~\ref{tab:summary} and level of income were considered as covariates in $tr(\cdot)$ and the countries were treated as the units in $tr_0(\cdot)$. The maximal number of splits considered was $S_{\text{max}} = 20$, and the optimal number of splits was selected based on the 10-fold cross-validation with the 1SE rule. The minimal bucket size was set to $n_{\text{mb}} = 100$ and the maximal depth of the tree to $d_{\text{max}} = 4$.  

\begin{figure}[t!]
\centering
\includegraphics[width = 0.7\textwidth]{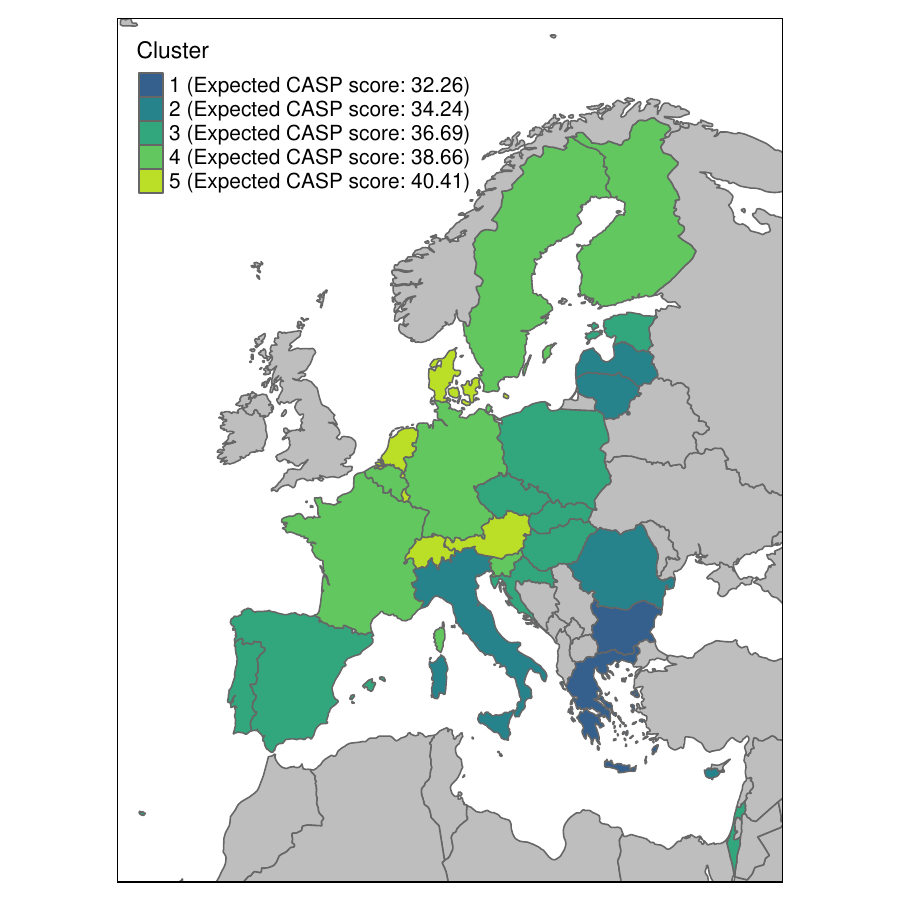}
\caption{Analysis of the SHARE data. Identified clusters of countries in $tr_0(\cdot)$ when fitting the TTSC model}\label{share_clusters}
\end{figure}

Figure~\ref{share_clusters} visualizes the results with regard to $tr_0(\cdot)$. Five clusters of countries were identified when fitting the model: The first cluster with the lowest expected CASP score is the smallest comprising only two countries (Bulgaria and Greece). The cluster with the second lowest expected QoL contains Eastern and Southern European countries (Cyprus, Italy, Latvia, Lithuania, and Romania). Central to Eastern European as well as the countries from the Iberian peninsula comprise the third cluster (Croatia, Czech Republic, Estonia, Hungary, Israel, Slovakia, Spain, Poland, and Portugal). The cluster with the second highest expected QoL is composed mostly of Central European as well as Scandinavian countries (Belgium, Finland, France, Germany, Slovenia, and Sweden). Finally, the cluster with the highest expected CASP score contains the five countries Austria, Denmark, Luxembourg, the Netherlands, and Switzerland.
These results indicate that the populations of wealthier countries tend to experience a higher QoL, which was shown previously in a study by~\citet{Diener1995} based on data from 101 nations. Specifically, Greece and Bulgaria, which constitute the cluster with lowest expected CASP score, exhibited the lowest gross domestic product (GDP) per capita of all countries in the EU in 2021, whereas Luxembourg, Denmark, the Netherlands, and Austria (i.e. EU countries in the fifth cluster) exhibited the highest, third, fourth and seventh highest GDP, respectively  \citep{Eurostat2024}. In addition, \citet{Niedzwiedz2014} analyzed data from wave 2 and 3 of SHARE and found that older adults from countries with more generous welfare regimes experienced higher QoL, which is confirmed by our findings: Scandinavian countries and countries with Bismarckian welfare regimes (e.g. Austria, France, Germany, and Switzerland) were placed in the the upper two clusters, while countries with Southern or Post-communist welfare regimes were mostly in clusters with lower expected QoL.   

Figure~\ref{share_subgroups} shows the results with regard to $tr(\cdot)$. Number of chronic diseases, level of income, and age of the individuals were selected as splitting variables during tree building and in total eleven different subgroups were identified. In particular, the number of chronic diseases was shown to have a very strong effect on QoL, as it was the first splitting variable in the root node and was selected for splitting most often. The corresponding results indicate that an increasing number of chronic diseases is associated with a decreased QoL, which aligns with the findings by \citet{Heyworth2009} and \citet{Rothrock2010}, who investigated the effect of chronic conditions on health-related QoL in United Kingdom (UK) and United States (US) citizens, respectively. Moreover, negative associations between the number of chronic diseases and QoL were frequently reported in the past decades  \citep{Marengoni2011, Makovski2019} and were also found in data from previous waves of SHARE \citep{Makovski2020, Rodriguez-Blazquez2020}. Moreover, household income is demonstrated to play an important role, where adults who are among the wealthier parts of the population of their respective country showed higher QoL. The positive effect of income on QoL was previously shown by \citet{Killingsworth2021} in US citizens and \citet{Knesebeck2007} in wave 1 of SHARE. Age was also selected for splitting but appeared to be only relevant for adults suffering from at least one chronic disease. 

\afterpage{
\begin{landscape}
\begin{figure}[htbp]
   \centering
    \includegraphics[height=0.7\textheight, trim = 6cm 0cm 0cm 2cm]{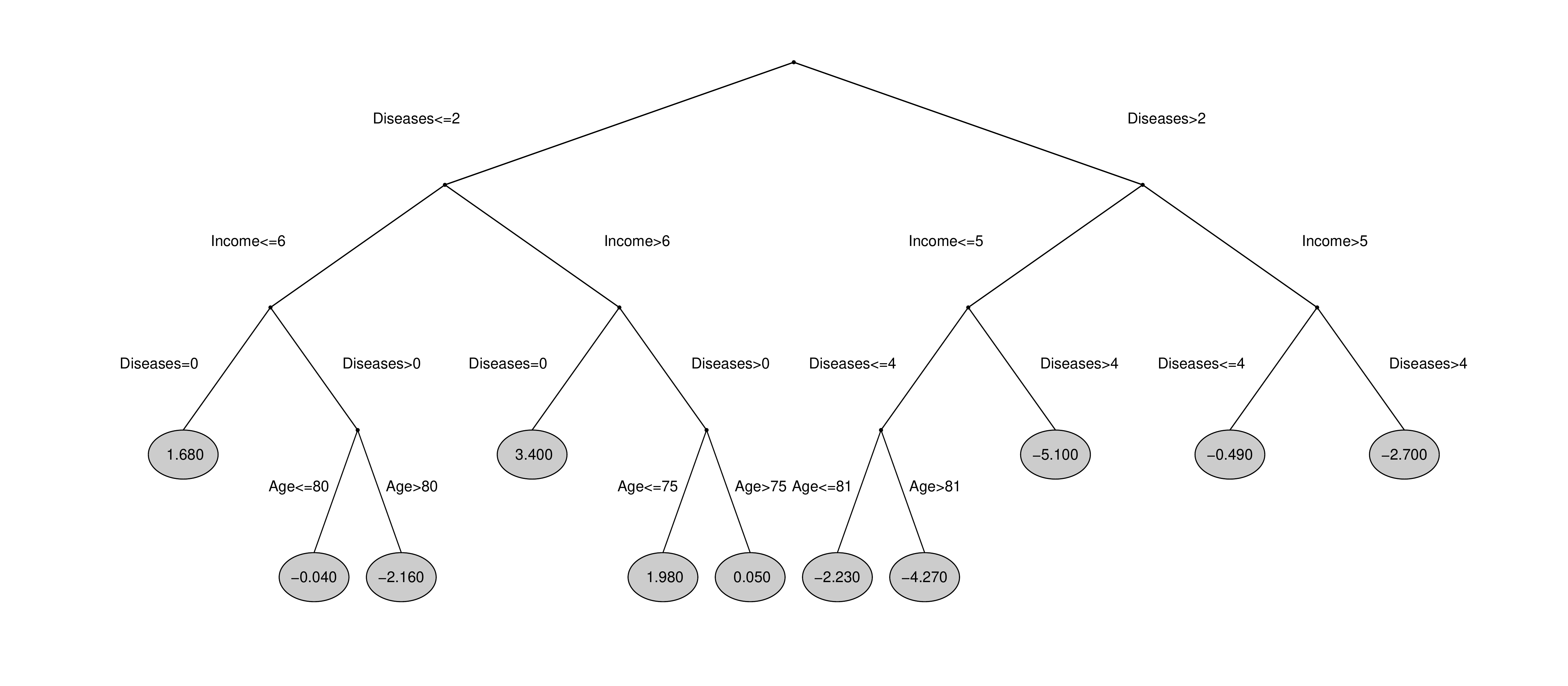}
    \caption{Analysis of the SHARE data. Identified subgroups of individuals in $tr(\cdot)$ when fitting the TTSC model. \textit{Diseases}, \textit{Income}, and \textit{Age} refer to the number of chronic diseases the person suffers from, the income decile the household falls in, and the age in years, respectively}
    \label{share_subgroups}
\end{figure}
\end{landscape} 
}

From Figure~\ref{share_subgroups} it is seen that the subgroup with the lowest expected QoL (among the people aged 50 years or older) constitutes individuals who suffer from more than four chronic diseases and are among the poorest 50 percent in terms of household income in their country. Individuals from this subgroup are expected to exhibit a by 5.10 points lower CASP score than the expected value of their country. On the other end, individuals with no chronic diseases that were among the wealthiest 40 percent of older adults from their country are shown to experience the highest QoL at a by 3.40 points increased CASP score compared to the expected value of their country. 


\section{Simulation study}
\label{sec:simulation}

To assess the performance and further analyze the properties of the proposed model, we considered different simulation scenarios. The simulation study was aimed to (i) investigate how the performance is affected by specific characteristics of the data, like the form of the data generating process (DGP; linear or tree-based), the number of units, the number of individuals per unit, and the number of covariates,  and (ii) to compare the proposed tree-structured model~\eqref{ttsc} to alternative models. 

\subsection{Simulation design}

We considered four simulation scenarios that were based on a DGP with predictor~\eqref{glmm} comprising linear effects of the covariates and random unit-specific intercepts (\textit{scenario 1}), a DPG with predictor~\eqref{reem} comprising tree-structured effects of the covariates and random unit-specific intercepts (\textit{scenario 2}), a DGP with predictor~\eqref{ltsc} comprising linear effects of the covariates and clustered fixed effects of the units (\textit{scenario 3}), and a DGP with predictor~\eqref{ttsc} composed of tree-structured effects of the covariates and clustered fixed effects of the units (\textit{scenario 4}). Further details on the DGPs will be given in the following subsections.

In all simulation scenarios, we considered six different settings and performed 100 replications each. In the first setting (\textit{setting~1}), we simulated data with $n=20$ units, $n_i=50$ observations per unit and $p=10$ potentially informative covariates. We included six metrically scaled covariates $X_{1}, \dots, X_{6}\sim N(0,1)$ and four binary covariates $X_{7},\dots,X_{10}\sim \text{Bin}(1, 0.5)$. Standard normally distributed error terms were included in the DGP. In the following we also refer to this first setting as \textit{base setting}. In the five other settings only one parameter compared to the base setting was modified to generate the data, while all the other parameters were kept fixed. In \textit{setting~2} and \textit{setting~3} we modified the ratio of units compared to the observations per units, setting $n=40/n_i = 25$ and $n=100/n_i = 10$, respectively. We considered a higher dimensional covariate space with $p=100$, $X_{11},\dots , X_{15} \sim N(0,1)$, and $X_{16},\dots ,X_{100}\sim \text{Bin}(1, 0.5)$  in \textit{setting~4}. In \textit{setting~5} the variance of the error terms was increased to $\sigma_{\varepsilon}^2=2$. The last setting (\textit{setting~6}) differs depending on the specific scenario and is described in the respective subsections.   

The following models were fitted to the simulated data in each scenario:

\begin{itemize}
\item[(i)] the linear mixed model~\eqref{glmm} with linear effects of the covariates and random unit-specific intercepts~(\textit{LMM}),
\item[(ii)] a LMM with variable selection by LASSO as proposed by \citet{Groll2014}, which applies an $L_1$-penalty on the linear effects of the covariates~(\textit{LMMP}),
\item[(iii)] the RE-EM tree~\eqref{reem} by \citet{Sela2012} with tree-structured effects of the covariates and random unit-specific intercepts~(\textit{RE-EM}),
\item[(iv)] the LMM tree by \citet{Fokkema2018}, which also has the form in Equation~\eqref{reem}, but compared to RE-EM applies the framework of conditional inference trees~(\textit{LMMT}),  
\item[(v)] the tree-structured FEM~\eqref{ltsc} by \citet{Berger2018} with linear effects of the covariates and tree-structured fixed effects of the units~(\textit{LTSC}),
\item[(vi)] a LTSC model, with variable selection applying backward selection~(\textit{LTSCB}),
\item[(vii)] the proposed tree-structured FEM~\eqref{ttsc} with tree-structured effects of the covariates and tree-structured fixed effects of the units~(\textit{TTSC}),
\item[(viii)] a model without any covariates and only a constant global intercept~(\textit{Null}), and
\item[(ix)] the true data-generating model~(\textit{Perfect}).
\end{itemize}

The LMMT model by \citet{Fokkema2018} implements a fitting procedure similar to the RE-EM tree, where the algorithm alternates between two steps: (i) Fitting the tree structure, while keeping the random effects fixed, and (ii) estimating the random effects, while keeping the tree structure fixed. Instead of the CART algorithm, the LMMT model applies conditional inference trees~\citep{Hothorn2006}. That is, in each iteration a test for parameter instability is carried out for each covariate and the covariate showing the strongest association with the outcome variable is selected for splitting (if it is significant at a predefined significance level~$\alpha$). The approach by \citet{Fokkema2018} is based on the framework of model-based recursive partitioning~\citep{Zeileis2008}, which additionally allows that in each terminal node of the tree a separate regression model is fitted. Here, we specified an intercept-only model to ensure comparability. Note that this is conceptually equivalent to the conditional inference-based version of the RE-EM tree by \citet{Fu2015}.

For the random intercepts in LMM, LMMP, RE-EM, and LMMT normality was assumed. The optimal penalty parameter $\lambda$ for the LASSO in LMMP was selected based on the Bayesian information criterion (BIC; \citealp{Schwarz1978}). The optimal number of splits in RE-EM, LTSC, LTSCB, and TTSC was selected based on 10-fold cross-validation with the 1SE rule (see also Section~\ref{sec:fitting}). The minimal bucket size (i.e.\@ the minimum number of observation required in a node) was set to $n_{\text{mb}} = \lfloor 0.1\cdot\sum_{i = 1}^{n}n_{i}\rfloor$ in all tree-based models (RE-EM, LMMT, LTSC, LTSCB, and TTSC). For the LTSCB model, first the LTSC model with the optimal number of splits was fitted and subsequently covariates were excluded from the linear predictor using backward selection based on BIC, while the tree structure of the unit-specific effects was kept fixed. Note that the perfect model cannot be fitted in practice as it is unknown and serves as reference, only.

\subsection{Evaluation criteria}

To assess the performance of the competing models in terms of goodness-of-fit, we considered the root mean squared error (RMSE) separately for the effects of the covariates and for the unit-specific effects. The RMSE of the covariate effects was calculated by
\begin{equation*}
\text{RMSE}_{\text{X}}= \sqrt{\frac{1}{n}\sum_{i = 1}^{n}\frac{1}{n_i}\sum_{j=1}^{n_{i}}\left(\tilde{\eta}_{\text{X}}(\bs{x}_{ij}) - \tilde{\hat{\eta}}_{\text{X}}(\bs{x}_{ij})\right)^2} \, ,
\end{equation*}
where $\tilde{\hat{\eta}}_{\text{X}}(\bs{x}_{ij}) = \hat{\eta}_{\text{X}}(\bs{x}_{ij}) - \frac{1}{n}\sum_{i'=1}^{n}\frac{1}{n_{i'}} \sum_{j' = 1}^{n_{i'}}\hat{\eta}_{\text{X}}(\bs{x}_{i'j'})$ corresponds to the covariate-specific deviation from the unit-specific expectation (also compare the adjustment of the coefficients in Section~\ref{subsec:TSC}). Specifically, for the models with linear effects of the covariates (LMM, LMMP, LTSC, and LTSCB) we have that $\hat{\eta}_{\text{X}}(\bs{x}_{ij}) = \hat{\beta}_1 x_{ij1} +\dots + \hat{\beta}_{p} x_{ijp}$ and for the models with tree-structured effects of the covariates (RE-EM, LMMT, and TTSC) we have that $\hat{\eta}_{\text{X}}(\bs{x}_{ij}) = \hat{tr}(\bs{x}_{ij})$. The RMSE of the unit-specific effects was calculated by 
\begin{equation*}
\text{RMSE}_{\text{I}}= \sqrt{\frac{1}{n}\sum_{i = 1}^{n}\frac{1}{n_i}\sum_{j=1}^{n_{i}}\left(\tilde{\eta}_{\text{I}}(i) - \tilde{\hat{\eta}}_{\text{I}}(i)\right)^2} \, ,
\end{equation*}
where $\tilde{\hat{\eta}}_{\text{I}}(i) = \hat{\eta}_{\text{I}}(i) + \frac{1}{n}\sum_{i'=1}^{n}\frac{1}{n_{i'}} \sum_{j' = 1}^{n_{i'}}\hat{\eta}_{\text{X}}(\bs{x}_{i'j'})$ corresponds to the expected outcome value of unit $i$. For models with random unit-specific intercepts (LMM, LMMP, RE-EM, and LMMT) this means $\hat{\eta}_{\text{I}}(i) = \hat{\beta}_{0} + b_{i}$ and for models with tree-structured fixed effects (LTSC, LTSCB, and TTSC) this means $\hat{\eta}_{\text{I}}(i) = \hat{tr}_{0}(i)$. Of note, for TTSC $\tilde{\eta}_{\text{X}}(\cdot)$ and $\tilde{\eta}_{\text{I}}(\cdot)$ could also directly be derived from the adjusted coefficients defined in~\eqref{adj}. The true values of $\tilde{\eta}_{\text{X}}(\cdot )$ and $\tilde{\eta}_{\text{I}}(\cdot )$ are determined analogously based on the true values.

In addition, true positive rates (TPR) and false positive rates (FPR) for the covariates were considered. The TPR is the proportion of informative covariates that were correctly identified to have an effect on the outcome variable and is given by
\begin{equation*}
\text{TPR}_{X} = \frac{1}{|\{k: \vartheta_k = 1\}|}\sum_{k:\vartheta_k = 1} I(\hat{\vartheta}_k = 1)\, ,
\end{equation*}
where $\vartheta =1$ if $X_k$ has an effect on the outcome variable and $\vartheta_k = 0$ otherwise. The FPR specifies the proportion of noise variables that were falsely identified to have an effect on the outcome variable. It is given by
\begin{equation*}
\text{FPR}_{X} = \frac{1}{|\{k: \vartheta_k = 0\}|}\sum_{k:\vartheta_k = 0} I(\hat{\vartheta}_k = 1)\, .
\end{equation*}

\subsection{Linear DGP with random unit-specific intercepts}

The first scenario was based on a DGP of the form
\begin{equation*}
y_{ij} = \beta_{1} x_{ij1} + \beta_{2} x_{ij2} + \beta_{7} x_{ij7} + b_{i} + \varepsilon_{ij}\,, 
\end{equation*}
with $\beta_{1}=0.8$, $\beta_{2} = 0.4$, and $\beta_{7} = 0.8$. Hence, three out of ten (or one hundred) covariates were informative. The random unit-specific intercepts $b_i$ follow a standard normal distribution. The data sets in settings 1 to 5 were generated as described above. In setting 6, we assumed that a correlation between $X_1$ as well as $X_2$ and the random intercepts $b_i$ is present. Specifically, a correlation of $\rho = 0.9$ was introduced by adopting the sequential procedure described in \citet{Oelker2017}. 

\begin{table*}[!h]
\caption{Results of the simulation study: Variable selection (scenario 1). Average true positive rates (TPR) and false positive rates (FPR) for the covariates in the six different settings. The table displays the results for all models that involve variable selection. Setting 1 serves as base setting with $n = 20$, $n_i =50$, $p = 10$ and error variance $\sigma_{\varepsilon}^2 = 1$}\label{tab:TPRFPR1}
\begin{center}
\begin{small}
\begin{tabularx}{\textwidth}{l l l Y Y Y Y Y Y}
\toprule
& Model & Setting & 1  & 2  & 3  & 4 & 5  & 6  \\
\hline
& & & \footnotesize{Base} & \footnotesize{$n= 40$} & \footnotesize{$n = 100$} & \footnotesize{$p = 100$} & \footnotesize{$\sigma_{\varepsilon}^2 = 2$} & \footnotesize{$\rho = 0.9$} \\
& & & & \footnotesize{$n_i = 25$} & \footnotesize{$n_i = 10$} & & & \\
\hline
TPR     & LMMP    & & 1.000   & 1.000   & 1.000   & 1.000   & 1.000   & 1.000   \\ 
        & RE-EM   & & 0.787   & 0.763   & 0.807   & 0.827   & 0.647   & 0.683   \\ 
        & LMMT    & & 0.963   & 0.937   & 0.963   & 1.000   & 0.967   & 0.843   \\ 
        & LTSCB   & & 1.000   & 1.000   & 1.000   & 1.000   & 1.000   & 1.000   \\ 
        & TTSC    & & 0.843   & 0.797   & 0.837   & 0.827   & 0.777   & 0.640   \\ \hline
FPR     & LMMP    & & 0.003   & 0.004   & 0.004   & 0.001   & 0.004   & 0.004   \\ 
        & RE-EM   & & 0.000   & 0.000   & 0.000   & 0.000   & 0.000   & 0.000   \\ 
        & LMMT    & & 0.000   & 0.001   & 0.000   & 0.003   & 0.007   & 0.004   \\ 
        & LTSCB   & & 0.003   & 0.004   & 0.003   & 0.008   & 0.003   & 0.000   \\ 
        & TTSC    & & 0.000   & 0.000   & 0.000   & 0.000   & 0.000   & 0.000   \\ 
\bottomrule
\end{tabularx}
\end{small}
\end{center}
\end{table*}
      
\begin{figure}[!h]
\centering
\includegraphics[width = 0.8\textwidth]{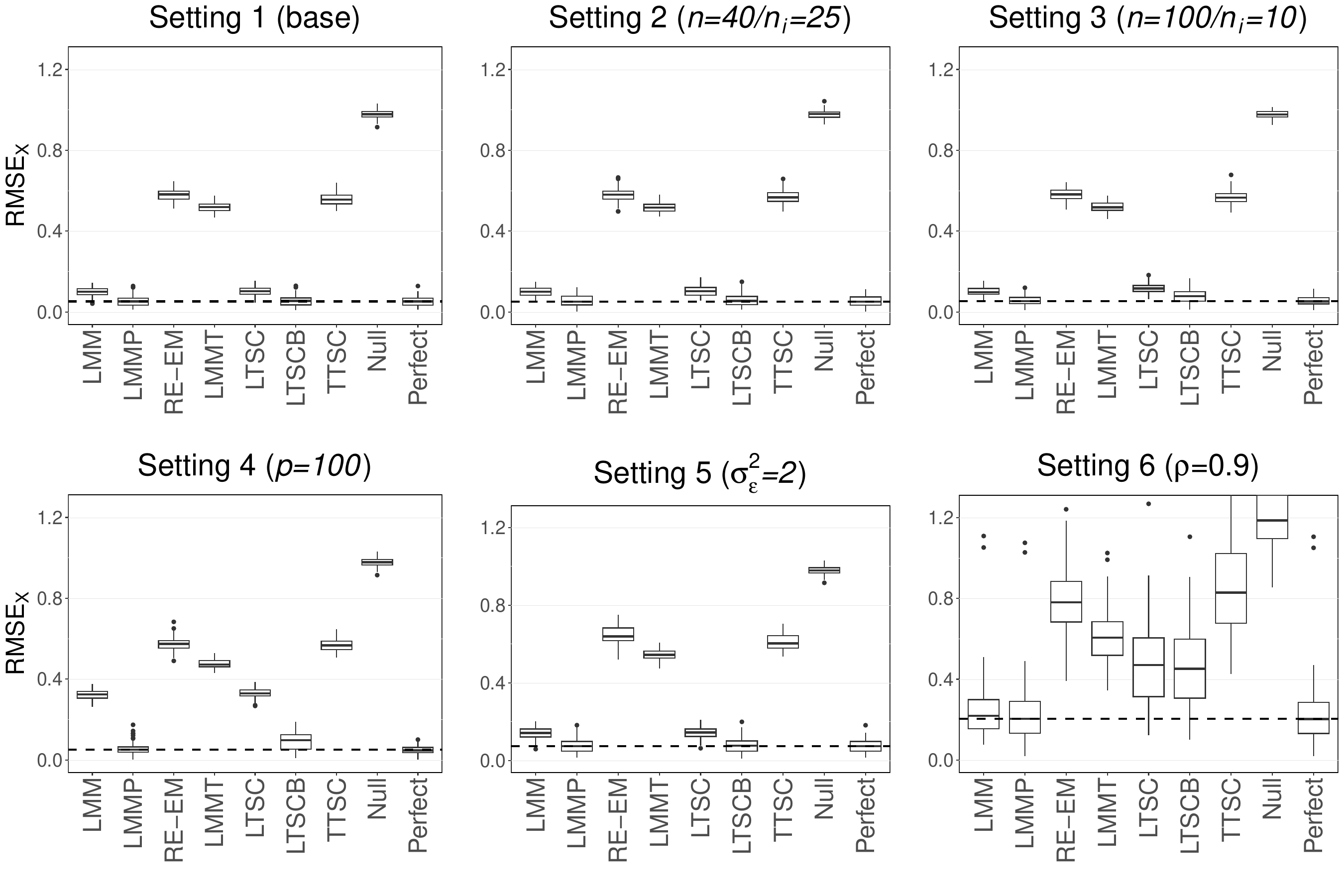}
\caption{Results of the simulation study:  $\text{RMSE}_{\text{X}}$  (scenario 1). Boxplots of the $\text{RMSE}_{\text{X}}$ in the six different settings. Setting 1 serves as base setting with $n = 20$, $n_i =50$, $p = 10$ and error variance $\sigma_{\varepsilon}^2 = 1$. The median values of the perfect model are marked by the dashed lines}\label{RMSEX1}
\end{figure}

The results in Table~\ref{tab:TPRFPR1} indicate that all considered models were very efficient in detecting the informative covariates. The models with linear effects (LMMP and LTSCB) exhibit perfect TPRs equal to one and very low FPRs below 0.01 across all settings. Among the models with tree-structured effects LMMT yielded the highest TPRs. RE-EM and TTSC showed more conservative results, which may be due to the application of the 1SE rule. Changing the ratio of $n$ to $n_i$ (settings 2 and 3) and increasing the number of noise variables (setting 4) only had a minor impact on the variable selection rates. In settings 5 and 6, however, it is seen that the TPRs decreased for the tree-structured models indicating that variable selection becomes less reliable for these methods if the error variance is large or informative covariates are strongly correlated with the random intercepts. 
These patterns can also be observed in Figure~\ref{RMSEX1}, which depicts the results for $\text{RMSE}_{\text{X}}$. The models with linear effects (LMMP and LTSCB), which follow the true DGP, are shown to perform best and even similarly well to the perfect model. The corresponding models without variable selection (LMM and LTSC) performed only slightly worse throughout all settings and consistently better than the tree-structured models. In addition, all of the considered models yielded a much higher variance in $\text{RMSE}_{\text{X}}$ if correlation between the informative covariates and the random intercepts was present (setting 6).

\begin{figure}[!h]
\centering
\includegraphics[width = 0.8\textwidth]{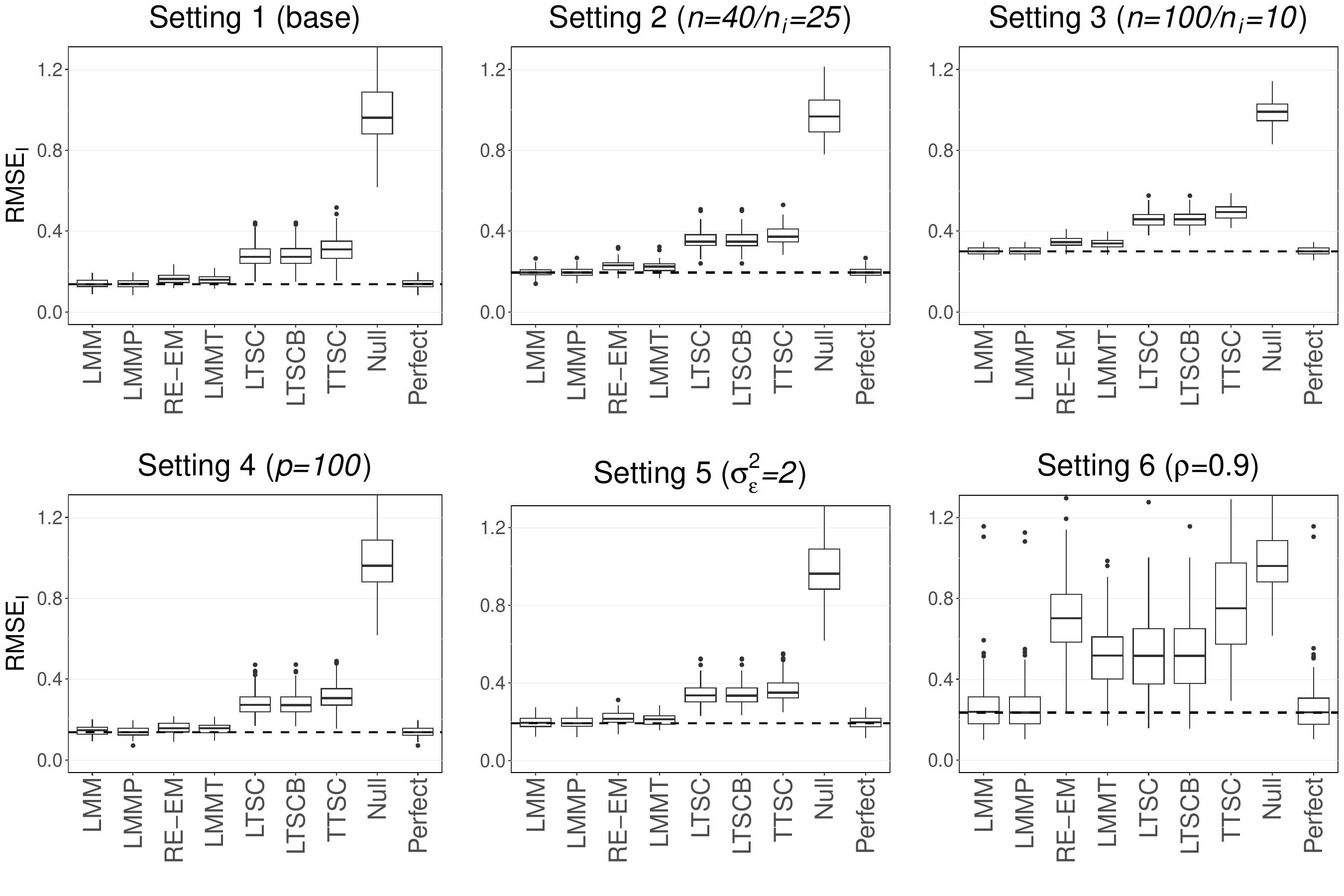}
\caption{Results of the simulation study:  $\text{RMSE}_{\text{I}}$  (scenario 1). Boxplots of the $\text{RMSE}_{\text{I}}$ in the six different settings. Setting 1 serves as base setting with $n = 20$, $n_i =50$, $p = 10$ and error variance $\sigma_{\varepsilon}^2 = 1$. The median values of the perfect model are marked by the dashed lines}\label{RMSEI1}
\end{figure}

Figure~\ref{RMSEI1} shows that the RMSE of the unit-specific effects were lowest for the models with random effects (LMM, LMMP, RE-EM, and LMMT) across all settings except for setting 6. Here, LMM and LMMP still performed best, but the tree-structured FEMs with linear covariate effects (LTSC and LTSCB) were beneficial compared to RE-EM and similar to LMMT. This is in line with the results obtained by~\citet{Berger2018} for correlated covariates. Further, it underlines that if correlation between the covariates and the random intercepts is present, a correct specification of the covariate effects (that structurally aligns with the DPG) is highly important for an unbiased estimation of the unit-specific effects. The proposed TTSC model exhibited the highest RMSEs compared to all other competitors except for the Null model, which was to be expected as it aligns the least with the structure of the DPG. 

\subsection{Tree-structured DGP with random unit-specific intercepts}

\begin{table*}[!t]
\caption{Results of the simulation study: Variable selection (scenario 2). Average true positive rates (TPR) and false positive rates (FPR) for the covariates in the six different settings. The table displays the results for all models that involve variable selection. Setting 1 serves as base setting with $n = 20$, $n_i =50$, $p = 10$ and error variance $\sigma_{\varepsilon}^2 = 1$}\label{tab:TPRFPR2}
\begin{center}
\begin{small}
\begin{tabularx}{\textwidth}{l l l Y Y Y Y Y Y}
\toprule
& Model & Setting & 1  & 2  & 3  & 4 & 5  & 6  \\
\hline
& & & \footnotesize{Base} & \footnotesize{$n= 40$} & \footnotesize{$n = 100$} & \footnotesize{$p = 100$} & \footnotesize{$\sigma_{\varepsilon}^2 = 2$} & \footnotesize{$\rho = 0.9$} \\
& & & & \footnotesize{$n_i = 25$} & \footnotesize{$n_i = 10$} & & & \\
\hline
TPR     & LMMP    & & 0.987   & 0.983   & 0.980   & 0.967   & 0.930   & 0.997   \\ 
        & RE-EM   & & 1.000   & 0.993   & 0.997   & 1.000   & 0.873   & 0.880   \\ 
        & LMMT    & & 1.000   & 1.000   & 1.000   & 1.000   & 1.000   & 0.997   \\ 
        & LTSCB   & & 0.950   & 0.940   & 0.943   & 0.987   & 0.833   & 0.930   \\ 
        & TTSC    & & 0.993   & 0.990   & 1.000   & 0.993   & 0.933   & 0.927   \\ \hline
FPR     & LMMP   & & 0.001   & 0.001   & 0.001   & 0.007   & 0.011   & 0.010   \\ 
        & RE-EM  & & 0.000   & 0.000   & 0.000   & 0.000   & 0.000   & 0.000   \\ 
        & LMMT   & & 0.001   & 0.001   & 0.002   & 0.001   & 0.019   & 0.004   \\ 
        & LTSCB  & & 0.001   & 0.000   & 0.001   & 0.007   & 0.004   & 0.004   \\ 
        & TTSC   & & 0.000   & 0.000   & 0.000   & 0.000   & 0.000   & 0.000   \\
\bottomrule
\end{tabularx}
\end{small}
\end{center}
\end{table*}

In the second scenario, the true DGP had the form
\begin{align*}
y_{ij} = &\gamma_{1} I(x_{ij1} \leq 0 \land x_{ij2}\leq 0 ) + \gamma_2 I(x_{ij1} \leq 0 \land x_{ij2}> 0 )\, +\\
& \gamma_{3} I(x_{ij1} > 0 \land x_{ij7} = 0 ) + \gamma_4 I(x_{ij1} > 0 \land x_{ij7}= 1 ) + b_{i} + \varepsilon_{ij}
\end{align*}
with $\gamma_{1} = -1.35$, $\gamma_{2} = -0.45$, $\gamma_{3} = 0.45$, and $\gamma_{4}=1.35$. Hence, again three out of ten (or one hundred) covariates were informative. Analogously to scenario~1, the unit-specific intercepts were standard normally distributed and for setting~6 a correlation of $\rho=0.9$ between $X_1$ as well as $X_2$ and the random intercepts was introduced.

\begin{figure}
\centering
\includegraphics[width = 0.8\textwidth]{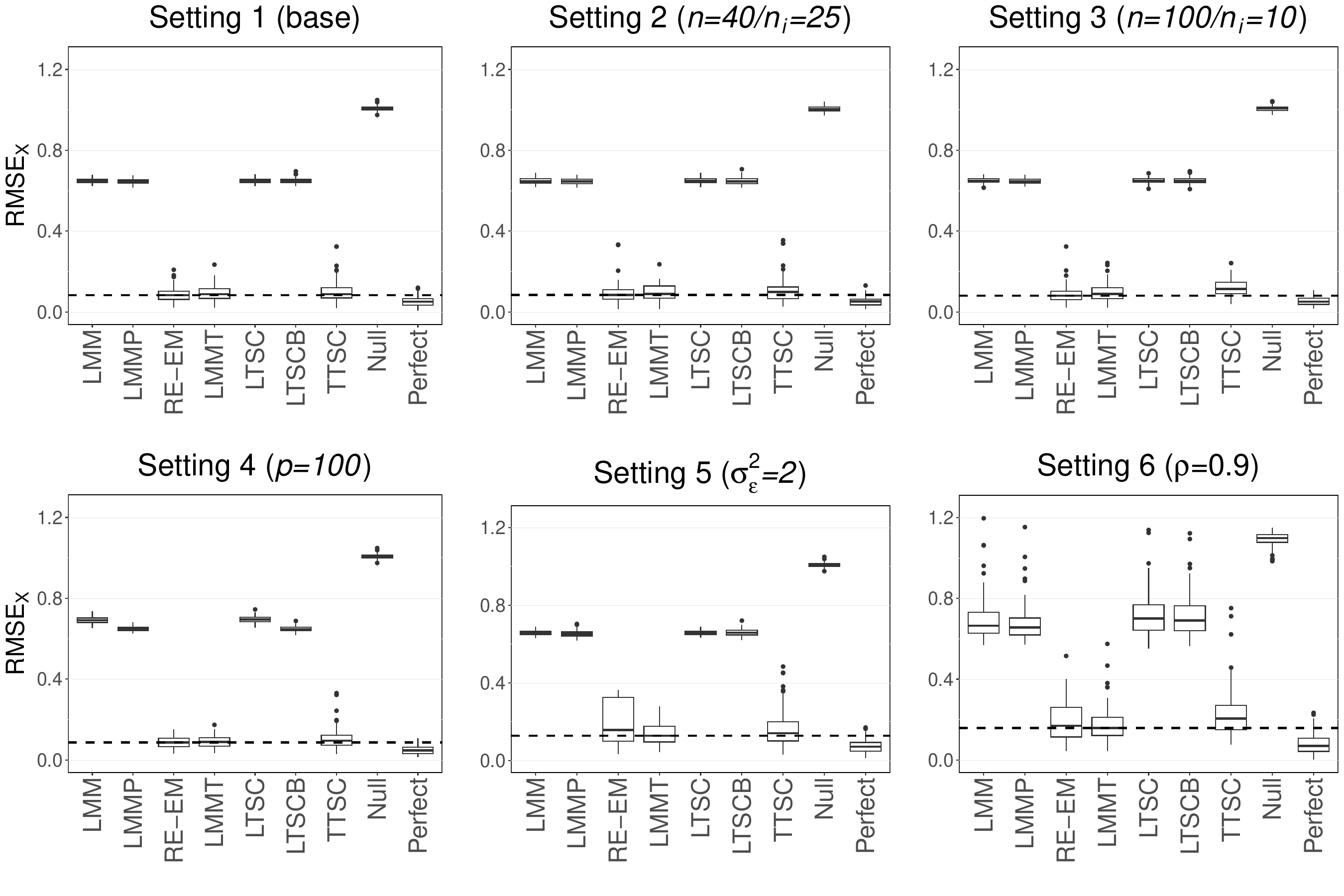}
\caption{Results of the simulation study:  $\text{RMSE}_{\text{X}}$  (scenario 2). Boxplots of the $\text{RMSE}_{\text{X}}$ in the six different settings. Setting 1 serves as base setting with $n = 20$, $n_i =50$, $p = 10$ and error variance $\sigma_{\varepsilon}^2 = 1$. The median values of the perfect model are marked by the dashed lines}\label{RMSEX2}
\end{figure}

It is seen from Table~\ref{tab:TPRFPR2} that the considered models exhibited high TPRs ($>0.8$) and low FPRs ($<0.02$) across all settings. The models with tree-structured effects of the covariates (RE-EM, LMMT, and TTSC) were superior to the other competitors with the highest TPRs and the lowest $\text{RMSE}_{\text{X}}$ close to the perfect model (see Figure~\ref{RMSEX2}). As in scenario 1, RE-EM and TTSC selected no non-informative covariates (FPRs $= 0$). Compared to LMMT, RE-EM and TTSC yielded slightly lower TPRs and higher RMSE values in setting 5 with higher error variance. Furthermore, the results in Table~\ref{tab:TPRFPR2} indicate the models with linear effects (LMMP and LTSC) were quite able to identify the informative covariates but the $\text{RMSE}_{\text{X}}$ demonstrate that they were unable to capture the non-linear covariate effects. Overall, the performance suffered in terms of variable selection and $\text{RMSE}_{\text{X}}$ if correlation between the informative covariates and the random intercepts occurred (setting 6).

\begin{figure}
\centering
\includegraphics[width = 0.8\textwidth]{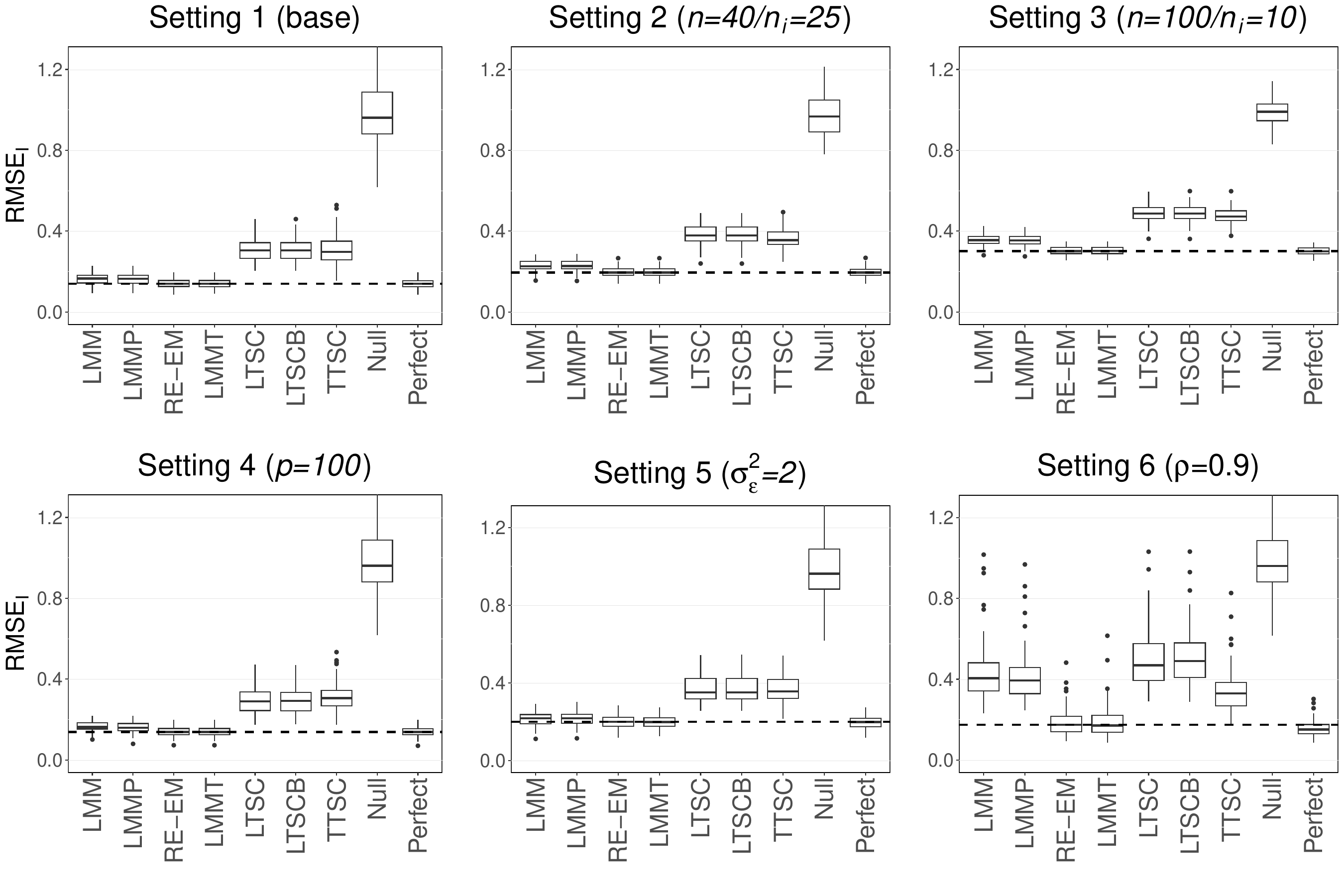}
\caption{Results of the simulation study:  $\text{RMSE}_{\text{I}}$  (scenario 2). Boxplots of the $\text{RMSE}_{\text{I}}$ in the six different settings. Setting 1 serves as base setting with $n = 20$, $n_i =50$, $p = 10$ and error variance $\sigma_{\varepsilon}^2 = 1$. The median values of the perfect model are marked by the dashed lines}\label{RMSEI2}
\end{figure}

Similar to scenario 1, the results in Figure~\ref{RMSEI2} show that the models with random effects (LMM, LMMP, RE-EM, and LMMT) were able to estimate the unit-specific effects more accurately than the tree-structured FEMs (LTSC, LTSCB, and TTSC) in settings 1 to 5. In setting 6, however, the models that capture tree-structured effects of the covariates showed superior performance compared to the linear models. In particular, TTSC yielded lower $\text{RMSE}_{\text{I}}$ than LMM and LMMP. The results in scenario 2 again indicate that neither the ratio of $n$ to $n_i$, the number of noise variables nor the variance of the error terms (varied in settings 2 to 5) changed the general pattern of the results. 
 
\subsection{Linear DGP with clustered unit-specific effects}

The data in the third scenario were generated according to the DGP
\begin{equation*}
y_{ij} = tr_0 (i) + \beta_1 x_{ij1} + \beta_{2}x_{ij2} + \beta_7 x_{ij7} + \varepsilon_{ij}
\end{equation*}
with $\beta_{1}=0.8$, $\beta_{2} = 0.4$, and $\beta_{7} = 0.8$. Hence, the linear predictor of the covariates coincided with scenario 1. In order to obtain clusters of units with the same intercepts, we drew a uniformly distributed random auxiliary variable $u_i\sim U(0,1)$ for each of the $n$ units. For settings 1 to 5, the effects of the units were then generated by
\begin{equation*}
tr_0 (i) =  \beta_{01} I\left(u_{i} \in \left[0, \frac{1}{3}\right]\right) + \beta_{02} I\left(u_{i} \in \left(\frac{1}{3}, \frac{2}{3}\right]\right) + \beta_{03}I\left(u_{i} \in \left(\frac{2}{3}, 1\right]\right)
\end{equation*}  
with $\beta_{01} = -1.25$, $\beta_{02} = 0$, and $\beta_{03} = 1.25$. That is, $C = 3$ clusters of units of roughly equal size were present in the data. In setting 6 we increased the number of clusters to $C = 6$ and applied the function
\begin{align*}
tr_0 (i) = & \beta_{01} I\left(u_{i} \in \left[0, \frac{1}{6}\right]\right) + \beta_{02} I\left(u_{i} \in \left(\frac{1}{6}, \frac{2}{6}\right]\right) + \beta_{03}I\left(u_{i} \in \left(\frac{2}{6}, \frac{3}{6}\right]\right) + \\
&\beta_{04} I\left(u_{i} \in \left(\frac{3}{6}, \frac{4}{6}\right]\right) + \beta_{05} I\left(u_{i} \in \left(\frac{4}{6}, \frac{5}{6}\right]\right) + \beta_{06}I\left(u_{i} \in \left(\frac{5}{6}, 1\right]\right)
\end{align*}
with $\beta_{01} = -1.5$, $\beta_{02} = -0.9$, $\beta_{03} = -0.3$, $\beta_{04} = 0.3$, $\beta_{05} = 0.9$, and $\beta_{06} = 1.5$.
 
\begin{table*}[!t]
\caption{Results of the simulation study: Variable selection (scenario 3). Average true positive rates (TPR) and false positive rates (FPR) for the covariates in the six different settings. The table displays the results for all models that involve variable selection. Setting 1 serves as base setting with $n = 20$, $n_i =50$, $p = 10$, $\sigma_{\varepsilon}^2 = 1$ and number of clusters $C=3$}\label{tab:TPRFPR3}
\begin{center}
\begin{small}
\begin{tabularx}{\textwidth}{l l l Y Y Y Y Y Y}
\toprule
& Model & Setting & 1  & 2  & 3  & 4 & 5  & 6  \\
\hline
& & & \footnotesize{Base} & \footnotesize{$n= 40$} & \footnotesize{$n = 100$} & \footnotesize{$p = 100$} & \footnotesize{$\sigma_{\varepsilon}^2 = 2$} & \footnotesize{$C = 6$} \\
& & & & \footnotesize{$n_i = 25$} & \footnotesize{$n_i = 10$} & & & \\
\hline
TPR    & LMMP   & & 1.000   & 1.000   & 1.000   & 1.000   & 1.000   & 1.000   \\ 
        & RE-EM  & & 0.753   & 0.817   & 0.760   & 0.773   & 0.590   & 0.760   \\ 
        & LMMT    & & 0.963   & 0.993   & 0.950   & 1.000   & 0.963   & 0.970   \\ 
        & LTSCB   & & 1.000   & 1.000   & 1.000   & 1.000   & 1.000   & 1.000   \\ 
        & TTSC    & & 0.783   & 0.803   & 0.803   & 0.807   & 0.710   & 0.817   \\ \hline
FPR     & LMMP  & & 0.006   & 0.004   & 0.009   & 0.004   & 0.006   & 0.006   \\ 
        & RE-EM  & & 0.000   & 0.000   & 0.000   & 0.000   & 0.000   & 0.000   \\ 
        & LMMT   & & 0.000   & 0.000   & 0.000   & 0.002   & 0.007   & 0.000   \\ 
        & LTSCB  & & 0.004   & 0.001   & 0.004   & 0.006   & 0.003   & 0.004   \\ 
        & TTSC   & & 0.000   & 0.000   & 0.000   & 0.000   & 0.000   & 0.000   \\ 
\bottomrule
\end{tabularx}
\end{small}
\end{center}
\end{table*}

\begin{figure}
\centering
\includegraphics[width = 0.8\textwidth]{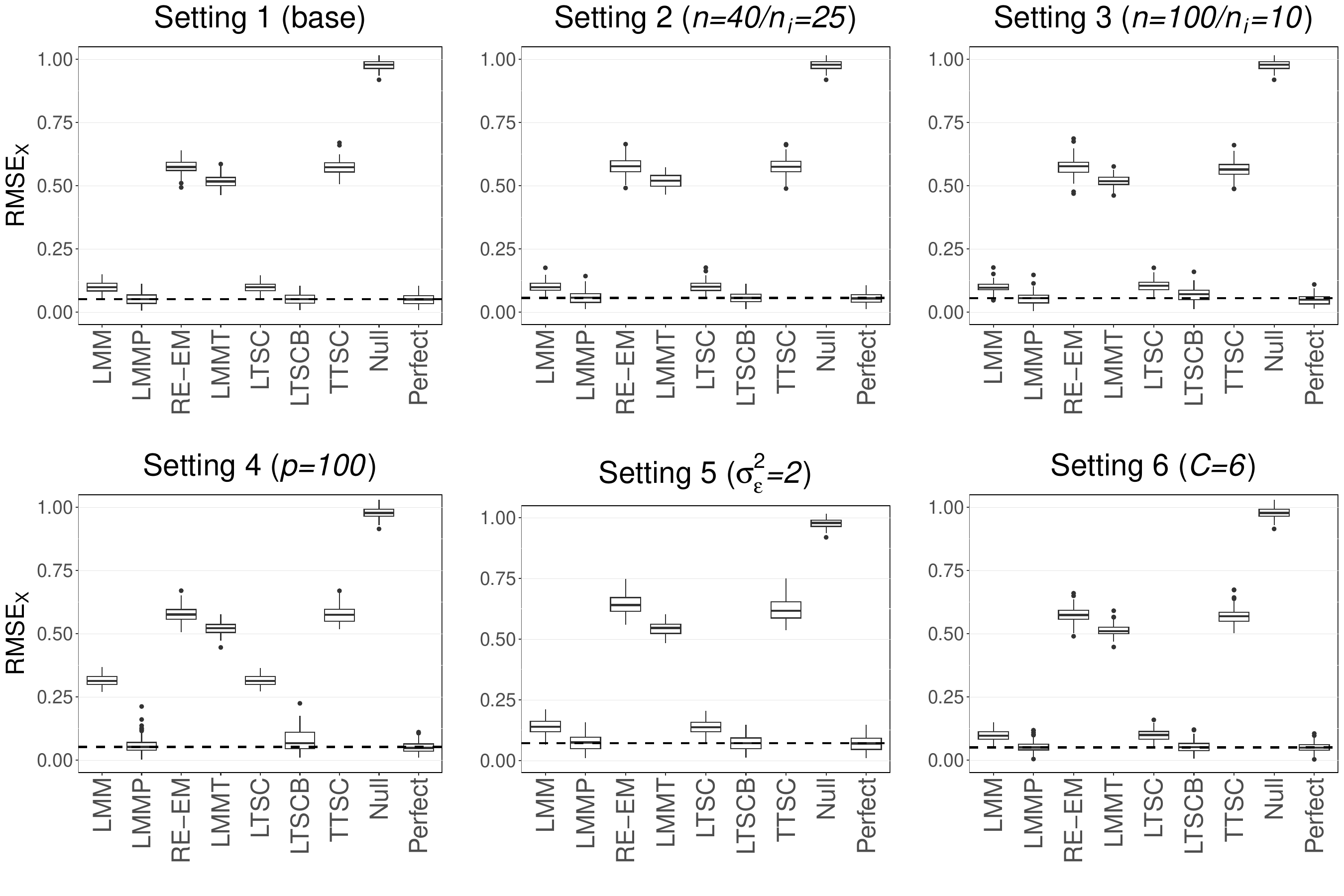}
\caption{Results of the simulation study:  $\text{RMSE}_{\text{X}}$  (scenario 3). Boxplots of the $\text{RMSE}_{\text{X}}$ in the six different settings. Setting 1 serves as base setting with $n = 20$, $n_i =50$, $p = 10$, $\sigma_{\varepsilon}^2 = 1$ and number of clusters $C=3$. The median values of the perfect model are marked by the dashed lines}\label{RMSEX3}
\end{figure}

The results shown in Table~\ref{tab:TPRFPR3} and Figure~\ref{RMSEX3} are fully in line with those of scenario~1 (see Table~\ref{tab:TPRFPR1} and Figure~\ref{RMSEX1}), where the effects of the covariates also followed a linear DGP. Specifically, the linear models (LMMP and LTSCB) showed perfect TPRs with low FPRs across all settings. In addition, LMMT performed best among the tree-structured models  and achieved TPRs of 0.95 or higher and decent $\text{RMSE}_{\text{X}}$ in all settings. Overall, non of the considered models showed considerable differences in variable selection rates and $\text{RMSE}_{\text{X}}$ compared to the base setting. Exceptions were the $\text{RMSE}_{\text{X}}$ of the models without variable selection (LMM and LTSC) in setting 4 and RE-EM and TTSC in setting 5. RE-EM, in particular, strongly deteriorated in terms of TPR as the variance of the error terms increased.    

\begin{figure}
\centering
\includegraphics[width = 0.8\textwidth]{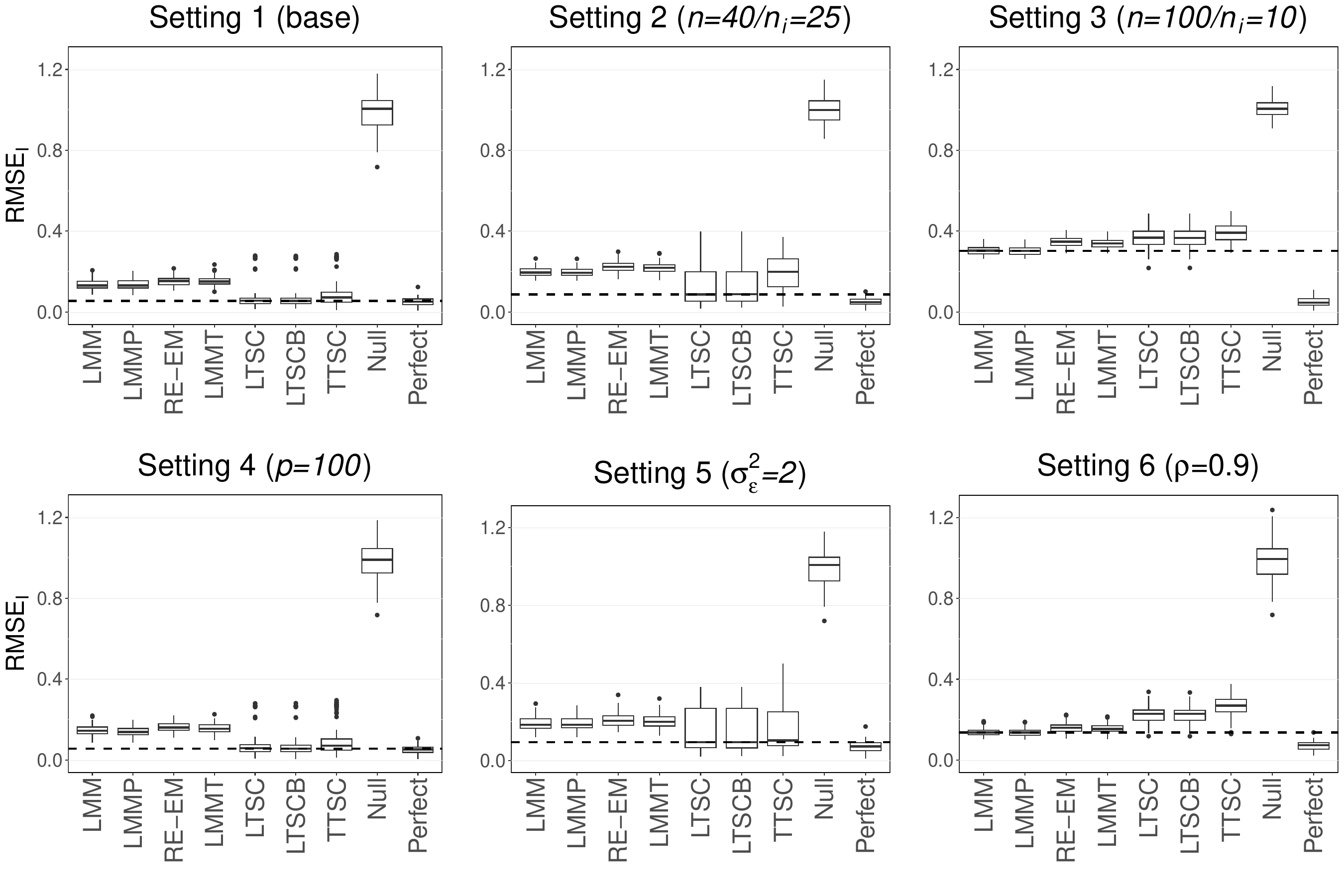}
\caption{Results of the simulation study:  $\text{RMSE}_{\text{I}}$  (scenario 3). Boxplots of the $\text{RMSE}_{\text{I}}$ in the six different settings. Setting 1 serves as base setting with $n = 20$, $n_i =50$, $p = 10$, $\sigma_{\varepsilon}^2 = 1$ and number of clusters $C=3$. The median values of the perfect model are marked by the dashed lines}\label{RMSEI3}
\end{figure}

Figure~\ref{RMSEI3}, which depicts the results of $\text{RMSE}_{\text{I}}$, shows that the tree-structured FEMs (LTSC, LTSCB, and TTSC) yielded more accurate estimates of the unit-specific effects than the models with random effects (LMM, LMMP, RE-EM, and LMMT) in settings~1,~4 and~5. As the number of units increased and the number of observations per unit decreased (settings~2 and~3), however, assuming random effects tended to be beneficial. In addition, a larger number of clusters also led to a higher $\text{RMSE}_{\text{I}}$ of the tree-structured FEMs compared to the models with random effects (setting~6). As the error variance was increased in setting 5, the tree-structured FEMs were still superior to the models with random effects, but showed much higher variability. Although the effects of the covariates followed a linear DGP, the TTSC model was not inferior in terms of $\text{RMSE}_{\text{I}}$ compared to LTSC and LTSCB.  

\subsection{Tree-structured DGP with clustered unit-specific effects}

In the fourth scenario, the true DGP had the form
\begin{align*}
y_{ij} = &tr_0(i) + \gamma_{1} I(x_{ij1} \leq 0 \land x_{ij2}\leq 0 ) + \gamma_2 I(x_{ij1} \leq 0 \land x_{ij2}> 0 )\, +\\
& \gamma_{3} I(x_{ij1} > 0 \land x_{ij7} = 0 ) + \gamma_4 I(x_{ij1} > 0 \land x_{ij7}= 1 ) + \varepsilon_{ij} \, ,
\end{align*}
where $\gamma_{1} = -1.35$, $\gamma_{2} = -0.45$, $\gamma_{3} = 0.45$, $\gamma_{4}=1.35$. Hence, the function $tr(\cdot)$ of the covariates coincided with scenario 2. The function $tr_0(\cdot)$ of the intercepts was defined analogously to scenario 3.

\begin{table*}[!h]
\caption{Results of the simulation study: Variable selection (scenario 4). Average true positive rates (TPR) and false positive rates (FPR) for the covariates in the six different settings. The table displays the results for all models that involve variable selection. Setting 1 serves as base setting with $n = 20$, $n_i =50$, $p = 10$, $\sigma_{\varepsilon}^2 = 1$ and number of clusters $C=3$}\label{tab:TPRFPR4}
\begin{center}
\begin{small}
\begin{tabularx}{\textwidth}{l l l Y Y Y Y Y Y}
\toprule
& Model & Setting & 1  & 2  & 3  & 4 & 5  & 6  \\
& & & \footnotesize{Base} & \footnotesize{$n= 40$} & \footnotesize{$n = 100$} & \footnotesize{$p = 100$} & \footnotesize{$\sigma_{\varepsilon}^2 = 2$} & \footnotesize{$C = 6$} \\
& & & & \footnotesize{$n_i = 25$} & \footnotesize{$n_i = 10$} & & & \\
\hline
TPR     & & LMMP    & 0.993   & 0.993   & 0.990   & 0.957   & 0.923   & 0.973   \\ 
       & & RE-EM   & 1.000   & 0.993   & 1.000   & 1.000   & 0.857   & 0.997   \\ 
       & & LMMT    & 1.000   & 1.000   & 1.000   & 1.000   & 1.000   & 1.000   \\ 
       & & LTSCB   & 0.953   & 0.923   & 0.913   & 1.000   & 0.867   & 0.930   \\ 
       & & TTSC    & 0.983   & 0.983   & 0.993   & 0.987   & 0.877   & 1.000   \\ \hline
FPR   &  & LMMP    & 0.017   & 0.017   & 0.016   & 0.005   & 0.023   & 0.013   \\ 
       & & RE-EM   & 0.000   & 0.000   & 0.000   & 0.000   & 0.000   & 0.000   \\ 
       & & LMMT    & 0.020   & 0.019   & 0.014   & 0.002   & 0.019   & 0.026   \\ 
       & & LTSCB   & 0.001   & 0.004   & 0.001   & 0.005   & 0.001   & 0.001   \\ 
       & & TTSC    & 0.000   & 0.000   & 0.000   & 0.000   & 0.001   & 0.000   \\ 
\bottomrule
\end{tabularx}
\end{small}
\end{center}
\end{table*}

\begin{figure}[!h]
\centering
\includegraphics[width = 0.8\textwidth]{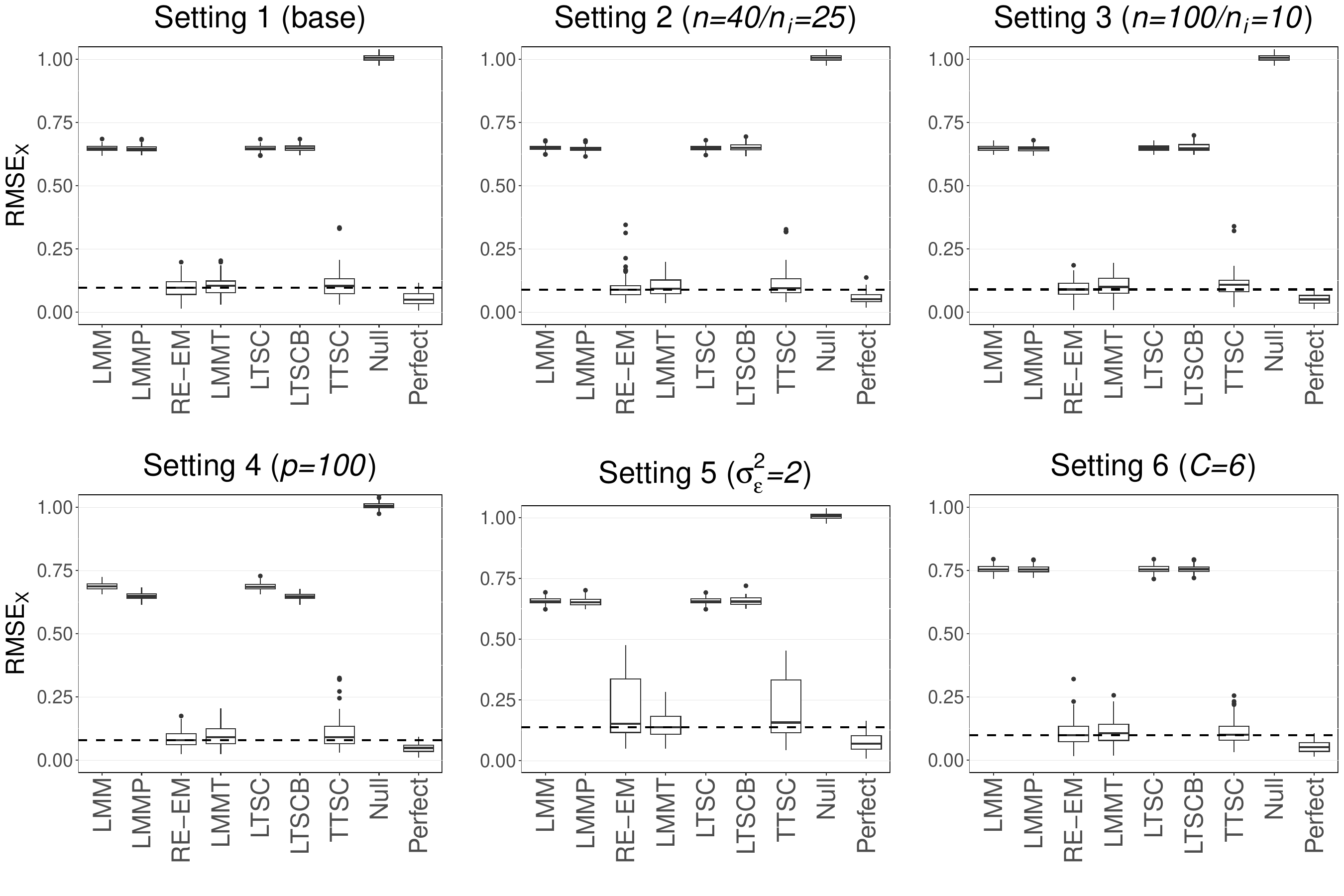}
\caption{Results of the simulation study:  $\text{RMSE}_{\text{X}}$  (scenario 4). Boxplots of the $\text{RMSE}_{\text{X}}$ in the six different settings. Setting 1 serves as base setting with $n = 20$, $n_i =50$, $p = 10$, $\sigma_{\varepsilon}^2 = 1$ and number of clusters $C=3$. The median values of the perfect model are marked by the dashed lines}\label{RMSEX4}
\end{figure}

The results in Table~\ref{tab:TPRFPR4} and Figure~\ref{RMSEX4} are comparable to the results in scenario~2 (see Table~\ref{tab:TPRFPR2} and Figure~\ref{RMSEX2}), where the effects of the covariates also followed a tree-structured DGP. It is seen that all the considered models were able to identify the informative covariates quite well, but the $\text{RMSE}_{\text{X}}$ values demonstrate that the linear models (LMM, LMMP, LTSC, and LTSCB) were unable to capture the non-linear covariate effects. The models with tree-structured effects of the covariates (RE-EM, LMMT and TTSC) yielded by far the lowest $\text{RMSE}_{\text{X}}$ across all settings. While the differences between the six different settings appear small, RE-EM and TTSC had worse performance in terms of TPR and $\text{RMSE}_{\text{X}}$ in setting 5 with increased variance of the error terms.

\begin{figure}[!h]
\centering
\includegraphics[width = 0.8\textwidth]{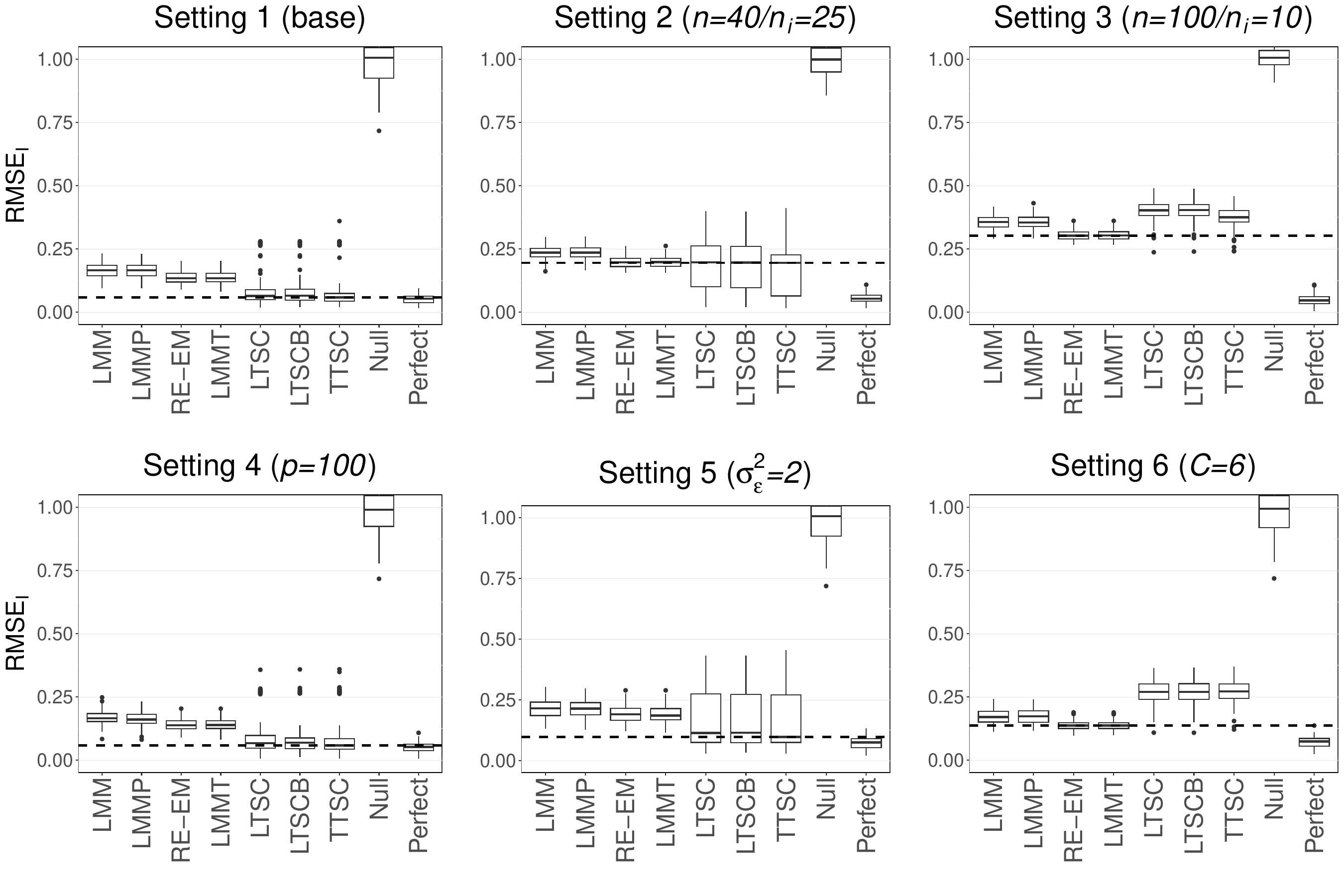}
\caption{Results of the simulation study:  $\text{RMSE}_{\text{I}}$  (scenario 4). Boxplots of the $\text{RMSE}_{\text{I}}$ in the six different settings. Setting 1 serves as base setting with $n = 20$, $n_i =50$, $p = 10$, $\sigma_{\varepsilon}^2 = 1$ and number of clusters $C=3$. The median values of the perfect model are marked by the dashed lines}\label{RMSEI4}
\end{figure}

The $\text{RMSE}_{\text{I}}$ shown in Figure~\ref{RMSEI4} strongly coincide with the results observed in scenario 3 (see Figure~\ref{RMSEI4}), where we also considered a DGP with clustered unit-specific effects. Specifically, the results indicate that the tree-structured FEMs (LTSC, LTSCB, and TTSC) were beneficial if the number of units was low, the number of observations per unit was high, and there were only few clusters of units present in the data. Overall, the TTSC model was shown to perform well in these settings in terms of variable selection and MRSE on the covariate- as well as the unit-level.  

To summarize the results of simulation scenarios 1 to 4, we made the following empirical key observations: 
\begin{enumerate}
\item All competitors showed high performance with regard to variable selection independent of the DPG. 
\item Based on the RMSE, the tree-structured models were able to capture non-linear effects and interactions well. 
\item Misspecification of unit-specific effects barely affects the goodness-of-fit of covariate effects. 
\item Misspecification of covariate effects leads to biased unit-specific effect estimates, if correlation between the covariates and the unit-specific effects is present. 
\item Tree-structured clustering is beneficial if the ratio of units to the observations per unit $n/n_i$ is low and the number of clusters of units $C$ is small.
\end{enumerate}

\section{Summary and discussion}
\label{sec:disc}

In order to analyze QoL in the group of elderly Europeans using data of SHARE, we developed a tailored tree-structured approach. Established methods for modeling clustered data allow to combine tree-structured effects of individual-level covariates with random country-specific effects \citep{Hajjem2011, Sela2012, Fu2015, Fokkema2018}, and to combine linear effects of covariates with clustered fixed country-specific effects \citep{Berger2018}. 
A method that simultaneously includes tree-structured effects of covariates and clustered fixed country-specific effects has not been available so far. In the present paper, we fill this gap. Specifically, the proposed model extends upon tree-structured clustering, which is designed for sparse modeling of unit-specific intercepts \citep{Berger2018}. We combine the tree representing unit-specific effects with a tree structure capturing effects of individual-level covariates. This second tree identifies subgroups of individuals that differ with regard to their outcome (the CASP score in SHARE). This accounts for non-linear effects and interactions between covariates, inherently performs variable selection and enables an accessible interpretation of parameters (see also the last paragraph in Section~\ref{subsec:TSC}). 

Our simulation study demonstrates that the proposed approach is competitive with alternative random effects-based approaches. Specifically, the proposed tree-structured FEM was shown to be advantageous if interactions between covariates were present and if there were only a few clusters of units with the same effect on the outcome. While random effects were also shown to work well in most settings and to be rather robust against violations of normality, confirming the findings in previous research (see, for example, \citealp{Bell2018}, and \citealp{Schielzeth2020}), the proposed tree-structured FEM yielded superior results in cases, where the number of units was low and the number of observations per unit was high. This is the case in SHARE with data from 28 countries and up to 3,000 observations per country. The analysis of SHARE presented in Section~\ref{sec:app} highlights the applicability of the proposed method and confirms important findings about QoL in older adults. 

While the focus in the simulation study and the application was on normally-distributed outcome variables, the proposed likelihood-based algorithm is generally applicable to differently scaled outcomes (including binary and discrete outcome variables). In addition, the predictor function is easily generalizable to an additive model of the form
\begin{equation}
\label{lttsc}
\eta(\bs{x}_{ij}, \bs{z}_{ij}) =  tr_{0}(i) + tr(\bs{x}_{ij})\, + \bs{\beta}^\top \bs{z}_{ij}\, ,
\end{equation}
where $\bs{z}_{ij}= (z_{ij1},\dots, z_{ijq})$ denotes an additional set of covariates with linear effects on the outcome. A random effects-based approach for modeling clustered data that also enables the combination of tree-structured and linear effects of the covariates was proposed by \citet{Gottard2023}. Their model can be represented by an additive predictor with a linear term and three tree structures for unit-varying and unit-constant covariates as well as for interactions between unit-varying and unit-constant covariates. 

In this paper, we reduced our considerations to clustered unit-specific intercepts. The proposed tree-structured algorithm, however, would also allow for clustered unit-specific effects of covariates (analogously to random slopes in random effects models). Referring to the set of covariates $\bs{z}_{ij}$, the model in Equation~\eqref{ltsc} can be extended to 
\begin{equation*}
\eta(\bs{x}_{ij}, \bs{z}_{ij}) =  tr_{0}(i) + \sum_{r = 1}^{q}tr_r (i) z_{ijr}+ tr(\bs{x}_{ij})\, ,
\end{equation*}
where the functions $tr_{r}(\cdot )$ are defined analogously to $tr_{0}(\cdot )$ as
\begin{equation*}
tr_{r} (i) = \sum_{\ell=1}^{C_r} \beta_{r\ell} \, I(i\in N_{r\ell} )\, ,
\end{equation*}
where $N_{r\ell}$ denotes the $\ell$-th cluster of the units with respect to the effect of $Z_{r}$ and $\beta_{r\ell}$ denotes the respective slope parameter. The fitting procedure described in Section~\ref{sec:fitting} can easily be adapted to this case by considering the possible splits in all $q+2$ trees in each step of the tree-building algorithm. In the first step, an order of the units $i \in \{1,\hdots,n\}$ needs to be determined with respect to each covariate, which is not necessarily the same.  
  
If the focus is on predictive performance, the proposed model can be extended to an ensemble method. In this vein, \citet{Adler2011} investigated bootstrap-based strategies for dealing with longitudinal data in random forests, and \citet{Hajjem2012} proposed a random effects-based random forest approach for modeling clustered data.

\section*{Acknowledgements}

\begin{sloppypar}
This paper uses data from SHARE Wave 9  (DOI: 10.6103/SHARE.w9ca900) see \citet{BoerschSupan2013} for methodological details.
The SHARE data collection has been funded by the European Commission, DG RTD through FP5 (QLK6-CT-2001-00360), FP6 (SHARE-I3: RII-CT-2006-062193, COMPARE: CIT5-CT-2005-028857, SHARELIFE: CIT4-CT-2006-028812), FP7 (SHARE-PREP: GA N$^{\circ}$ 211909, SHARE-LEAP: GA N$^{\circ}$ 227822, SHARE M4: GA N$^{\circ}$ 261982, DASISH: GA N$^{\circ}$ 283646) and Horizon 2020 (SHARE-DEV3: GA N$^{\circ}$ 676536, SHARE-COHESION: GA N$^{\circ}$ 870628, SERISS: GA N$^{\circ}$ 654221, SSHOC: GA N$^{\circ}$ 823782, SHARE-COVID19: GA N$^{\circ}$ 101015924) and by DG Employment, Social Affairs \& Inclusion through VS 2015/0195, VS 2016/0135, VS 2018/0285, VS 2019/0332, VS 2020/0313, SHARE-EUCOV: GA N$^{\circ}$101052589 and EUCOVII: GA N$^{\circ}$101102412. Additional funding from the German Federal Ministry of Education and Research (01UW1301, 01UW1801, 01UW2202), the Max Planck Society for the Advancement of Science, the U.S. National Institute on Aging (U01\_AG09740-13S2, P01\_AG005842, P01\_AG08291, P30\_AG12815, R21\_AG025169, Y1-AG-4553-01, IAG\_BSR06-11, OGHA\_04-064, BSR12-04, R01\_AG052527-02, R01\_AG056329-02, R01\_AG063944, HHSN271201300071C, RAG052527A) and from various national funding sources is gratefully acknowledged (see www.share-eric.eu).
\end{sloppypar}

\section*{Statements and declarations}

\textbf{Competing interests:} The authors have no competing interests to declare.

\bibliography{bib_doi_new}

\end{document}